\documentclass[reprint,twocolumn,showpacs,amsmath,amssymb,aps, prb ,superscriptaddress]{revtex4-1}

\usepackage{latexsym}

\usepackage[dvipdfmx]{graphicx}
\usepackage{color}
\usepackage{comment}
\usepackage[dvipdfmx]{graphicx}
\usepackage{bm}
\usepackage{multirow}
\usepackage{epstopdf}
\usepackage{subcaption}
\newcommand{\sidecaption}[1]
{\raisebox{\abovecaptionskip}{\begin{subfigure}[t]{1.6em}
  \caption[singlelinecheck=on]{}
  \label{#1}
\end{subfigure}}\ignorespaces}
\begin{document}

\title{Kinetic magnetoelectric effect in topological insulators}%

\author{Ken Osumi}%
\affiliation{Department of Physics, Tokyo Institute of Technology, Tokyo 152-8551, Japan}
\author{Tiantian Zhang}%
\affiliation{Department of Physics, Tokyo Institute of Technology, Tokyo 152-8551, Japan}
\affiliation{TIES, Tokyo Institute of Technology, Tokyo 152-8551, Japan}
\author{Shuichi Murakami}%
\email{murakami@stat.phys.titech.ac.jp}
\affiliation{Department of Physics, Tokyo Institute of Technology, Tokyo 152-8551, Japan}
\affiliation{TIES, Tokyo Institute of Technology, Tokyo 152-8551, Japan}
\altaffiliation{murakami@stat.phys.titech.ac.jp}
\date{\today}%

\begin{abstract}
\noindent{\bf Abstract.}
The kinetic magnetoelectric effect is an orbital analogue of the Edelstein effect and offers an additional degree of freedom to control magnetisation via the charge current. Here we theoretically propose a gigantic kinetic magnetoelectric effect in topological insulators and interpret the results in terms of topological surface currents. We construct a theory of the kinetic magnetoelectric effect for a surface Hamiltonian of a topological insulator, and show that it well describes the results by direct numerical calculation. This kinetic magnetoelectric effect depends on the details of the surface, meaning that it cannot be defined as a bulk quantity. We propose that Chern insulators and $Z_2$ topological insulators can be a platform with a large kinetic magnetoelectric effect, compared to metals by 5 - 8 orders of magnitude, because the current flows only along the surface. We demonstrate the presence of said effect in a topological insulator, identifying Cu$_2$ZnSnSe$_4$ as a potential candidate. 
\end{abstract}
\maketitle

\noindent{\bf Introduction.} In recent years, new responses leading to orbital magnetization have been proposed in systems without inversion symmetry \cite{yoda1,yoda2, Furukawa2017,zhong, Tsirkin2018, Wang2020, Moore2010, Sodemann2015, Ma2018, Shalygin2012, Koretsune2012, Furukawa2020, Rou2017, Sahin2018, Hara2020, He2020,PhysRevB.92.235205,PhysRevB.100.075136}. One of the focuses is conversion of electron current and magnetization on crystal structure with low symmetry. Among such proposals are kinetic magnetoelectric effect (KME) \cite{yoda1,yoda2, Furukawa2017, Hara2020, Shalygin2012, Koretsune2012, Sahin2018, Rou2017,PhysRevB.100.075136}, also called orbital Edelstein effect, i.e. current-induced orbital magnetization, and the gyrotropic magnetic effect \cite{zhong, Tsirkin2018, Wang2020}. These effects have similar response coefficients. In particular, KME is an orbital analog of the Edelstein effect \cite{Edelstein1990, Ivchenko1978, levitov1985nazarov}. KME emerges even in systems without spin-orbit interactions \cite{yoda1, Furukawa2017}. In particular, the KME emerges in crystals with a chiral structure \cite{yoda1,yoda2, Furukawa2017}, similar to the phenomenon in which the solenoid creates a magnetic field when a current flows. As a similar context, the recent finding of spin-selective electron transport through chiral molecules,
the so-called chirality-induced spin selectivity (CISS) effect\cite{Naaman2012, Gohler2011, Kettner2018, Bloom2016, Dor2014}, suggests an alternative method of using organic materials as spin filters for spintronics applications.

In the KME, the electric field induces the magnetization, which may look similar to the magnetoelectric (ME) effect\cite{Kimura2003, Katsura2005, Fiebig2005, Spaldin2005, Wilczek1987, Essin2009, Essin2010,Malashevich2010, Coh2011}. Nonetheless, in the spin Edelstein effect and KME, metallic systems are considered, and nonequilibrium electron distribution 
by the electric field is a key to generate magnetization.  
On the other hand, the ME effect originates from the change of the electronic band structure by the electric field, while
the electron distribution is assumed to stay in equilibrium. Thus 
the ME effect is mainly considered in insulators, but can also 
exist in metals \cite{PhysRevResearch.2.043060,PhysRevB.103.045401}.
In accordance with this differences in mechanisms between the KME and the ME effects, 
their symmetry requirements are different. In the ME effect, inversion and time-reversal symmetries should be broken. On the other hand, in the Edelstein effect, the inversion symmetry should be broken, {but the time-reversal symmetry can either be preserved or borken}. It is seen in chiral systems 
\cite{yoda1,yoda2, Furukawa2017, Shalygin2012, Koretsune2012, Sahin2018, Rou2017}, and in polar systems \cite{Hara2020}.

Magnetoelectric tensors may require careful consideration of boundaries. 
While orbital magnetization is independent on the boundary \cite{ceresoli, thonhauser, Xiao2005, Chen2012, bianco, Marrazzo2016}, the orbital magnetization when an electric field is applied may not have such properties. The general orbital magnetoelectric response \cite{Essin2009, Essin2010,Malashevich2010, Coh2011} depends on the boundary \cite{Chen2012}. Therefore, it is important to study the effect of the boundary of the response of orbital magnetization.

In this paper, we investigate KME in topological insulators such as three-dimensional Chern insulators and  $Z_2$ topological insulators ($Z_2$-TIs) \cite{Fu2007, Moore2007, Hsieh2008} in which the currents are localized on the surface. First, we calculate the KME in three-dimensional topological insulators with chiral crystal structure. Second, we derive the KME based on the surface Hamiltonian, and we show that this effect depends on surface states. Finally, we propose candidate materials for this effect and estimate the values of the KME. By comparing the results with the results in a chiral semiconductor tellurium \cite{Tsirkin2018}, we show that in topological insulators the orbital magnetization as a response to the current is much larger than metals by many orders of magnitude.

\begin{figure*}[ptb]
\raisebox{-\height}{\includegraphics[width=0.8\textwidth]{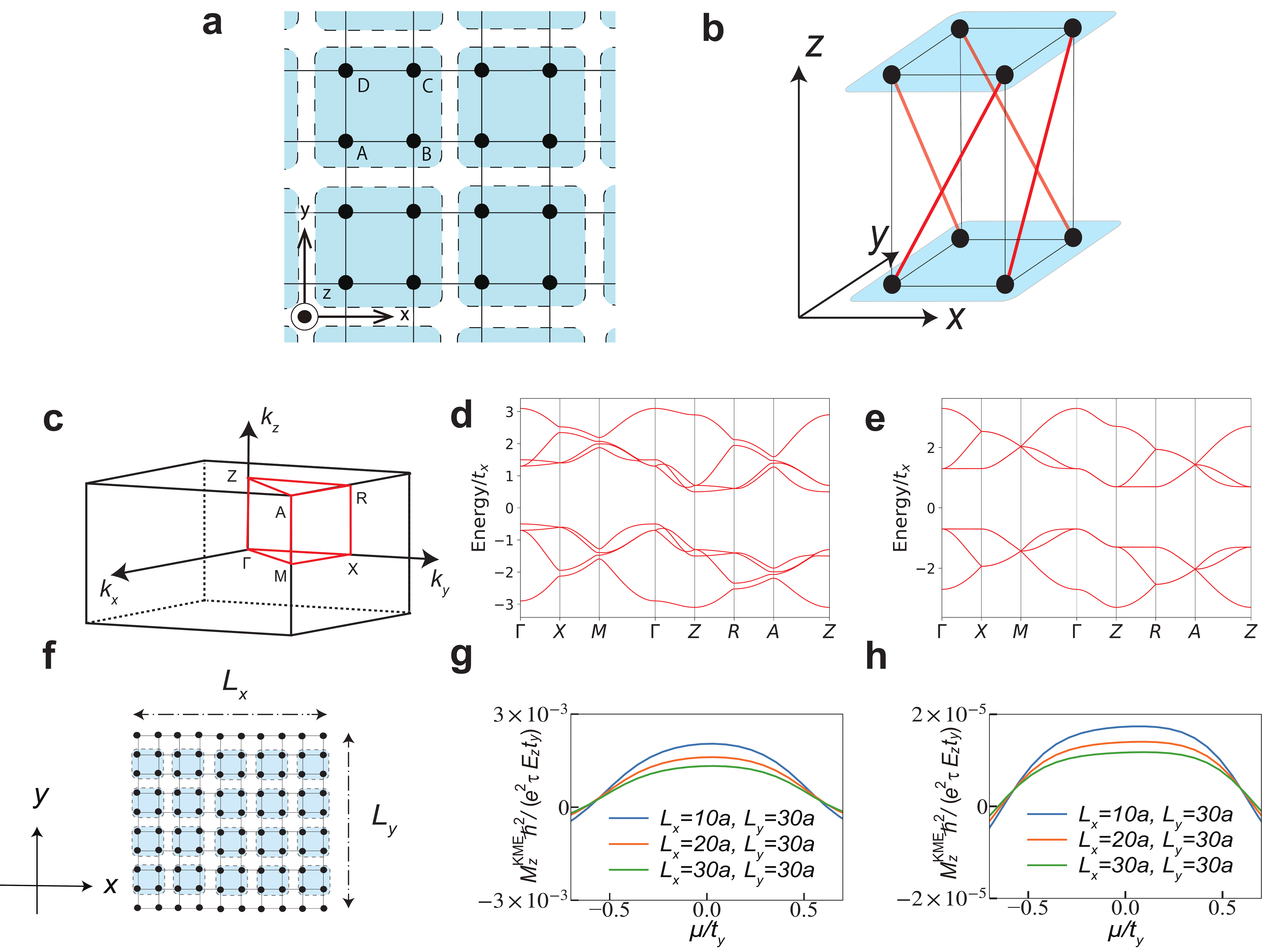}}
 \caption{{\bf The model of a Chern insulator with a chiral structure and model calculation of the kinetic magnetoelectric effect (KME).} ({\bf a}) Individual layer of the model forming a square lattice. The blue regions surrounded by the broken line are the unit cells consisting of four sublattices. ({\bf b}) Schematic picture of the chiral hopping (red) between the two neighboring layers. These hoppings form structure similar to right-handed solenoids. ({\bf c}) Brillouin zone of our model with high-symmetry points. ({\bf d, e}) Energy bands for the Hamiltonian $H$ with parameters ({\bf d}) $t_x=t_y=m=b_x=b_y$, $t_3=0.1t_y$ and $t_4=0.15t_y$ and ({\bf e})  $t_x=t_y=m=b_x=b_y$, $t_3=0.001t_y$ and $t_4=0.15t_y$. The symbols such as $\Gamma$ and $X$ represent high-symmetry points in the wavevector space. ({\bf f}) One-dimensional model with periodic boundary condition in $z$ direction. It has a rectangle shape 
 of a size $L_x\times L_y$ within the $xy$ plane. The dots are the sites, and th blue squares denote the unit cells, as specified in (a). In order to 
 see the boundary effect on KME, the outermost layers on the $xz$ surface has no chiral hoppings and those on the $yz$ surface has chiral hoppings. ({\bf g, h}) KME calculated with parameters ({\bf g})  $t_x=t_y=m=b_x=b_y$, $t_3=0.1t_y$ and $t_4=0.15t_y$ and ({\bf h})  $t_x=b_x=t_y$, $t_y=b_y=m$, $t_4=0.15t_y$, and $t_3=0.001t_y$. The blue, orange and green curves represent the data
 for $L_x=10a$, $L_x=20a$, and $L_x=30a$, respectively, while $L_y$ is fixed as $L_y=30a$.}
  \label{f1}
\end{figure*}

\noindent{\bf Results} \\
\noindent{\bf Formulation for KME.}
We consider a crystal in a shape of a cylinder along the $z$-axis, and calculate its orbital magnetization along the $z$-axis generated by the current along the $z$-axis. Let $c$ be the lattice constant along the $z$-axis. We introduce the velocity operator $\mathbf{v}$ as
\begin{math}
\mathbf{v}=-\frac{i}{\hbar}[\mathbf{r},H],
\label{v}
\end{math}
where $\mathbf{r}$ is the position operator and $H$ is the Hamiltonian.
In the limit of the system length along the $z$-axis to be infinity, the orbital magnetization at zero temperature is
\begin{eqnarray}
M_z&=&\frac{1}{2\pi}\int^{\pi/c}_{-\pi/c}dk_z\frac{1}{S}\sum_n^Nf(E_n(k_z))\nonumber\\
&&\times\left(-\frac{e}{2}\right)
\int dxdy \psi_{n,k_z}^{\dagger}(x,y)(\mathbf{r} \times \mathbf{v})_z\psi_{n,k_z}(x,y),
\label{eq:mz}
\end{eqnarray}
where
\begin{math}
-e
\end{math} 
is the electron charge,  
\begin{math}
\psi_{n,k_z}(x,y)
\end{math}
and
\begin{math}
E_n
\end{math}
are the $n$th occupied eigenstates and energy eigenvalues of $H$ at the Bloch wavenumber $k_z$, respectively.
\begin{math}
f(E)
\end{math}
is the distribution function at the energy $E$, $S$ is the cross section of the crystal along the $xy$-plane and $N$ is the number of occupied states.
{We note that the position
operator $\mathbf{r}$ is unbounded and problematic if a system is infinite in some directions. Nonetheless, in the present case, the system is infinite only along the $z$ direction,
while $(\mathbf{r} \times \mathbf{v})_z$ does not involve $z$, which means Eq.~(\ref{eq:mz}) is well defined. The unit of the orbital magnetization is [A/m] in the SI unit.}

Then, within the Boltzman approximation \cite{zhong}, the applied electric field
\begin{math} 
E_z
\end{math}
changes 
\begin{math}
f(E)
\end{math}
from
\begin{math}
f^0(E)
\end{math}
into
$f(E)=f^0(E)+\frac{e\tau E_z}{\hbar}\frac{\partial f^0(E)}{\partial k_z}
$
in a linear order in $E_z$, where
\begin{math}
\tau
\end{math}
is the relaxation time assumed to be constant and
\begin{math}
f^0(E)
\end{math}
is the Fermi distribution function
\begin{math}
f^0(E)=(e^{\beta(E-\mu)}+1)^{-1},
\end{math} 
\begin{math}
\beta=1/k_BT,
\end{math}
\begin{math}
k_B
\end{math}
is the Boltzman constant and
\begin{math}
\mu
\end{math}
is the chemical potential. Then the orbital magnetization is generated as
\begin{eqnarray}
\displaystyle
M_z^{\rm KME}&=&\frac{1}{2\pi}\int^{\pi/c}_{-\pi/c}dk_z\frac{1}{S}\sum_n^N\frac{e\tau E_z}{\hbar}\frac{\partial f^0(E_n(k_z))}{\partial k_z}\nonumber\\
&&\times \left(-\frac{e}{2}\right)
\int dxdy \psi_{n,k_z}^{\dagger}(x,y)(\mathbf{r} \times \mathbf{v})_z\psi_{n,k_z}(x,y)\rangle.
\label{3}
\end{eqnarray}
This is the KME. 
{In Chern insulators, because the TRS is broken, an orbital magnetization is nonzero in equilibrium, and a current leads to a change in an orbital magnetization due to the KME. On the other hand, in $Z_2$-TIs, 
an orbital magnetization vanishes in equilibirium due to TRS, and the KME leads to appearance of an orbital magnetization.}
{We note that in addition to the off-equilibrium electron distribution discussed above, the electric field also modifies the electronic states. In $Z_2$-TIs, where TRS is preserved,
this modulation of the electronic states does not lead to the orbital magnetization in the linear order in $E_z$, because the TRS is preserved within this order.
On the other hand, in topological insulators without TRS, such as Chern insulators, this may also contributes to the orbital magnetization. This mechanism is similar to the conventional ME effect in insulators, and has been discussed also in metals \cite{PhysRevResearch.2.043060,PhysRevB.103.045401}}

This calculation method is different from that in the previous study on metals \cite{yoda1,yoda2,zhong}, where bulk contribution in a system infinite along $x$ and $y$ directions are calculated. 

\begin{figure*}
\raisebox{-\height}{\includegraphics[width=0.7\textwidth]{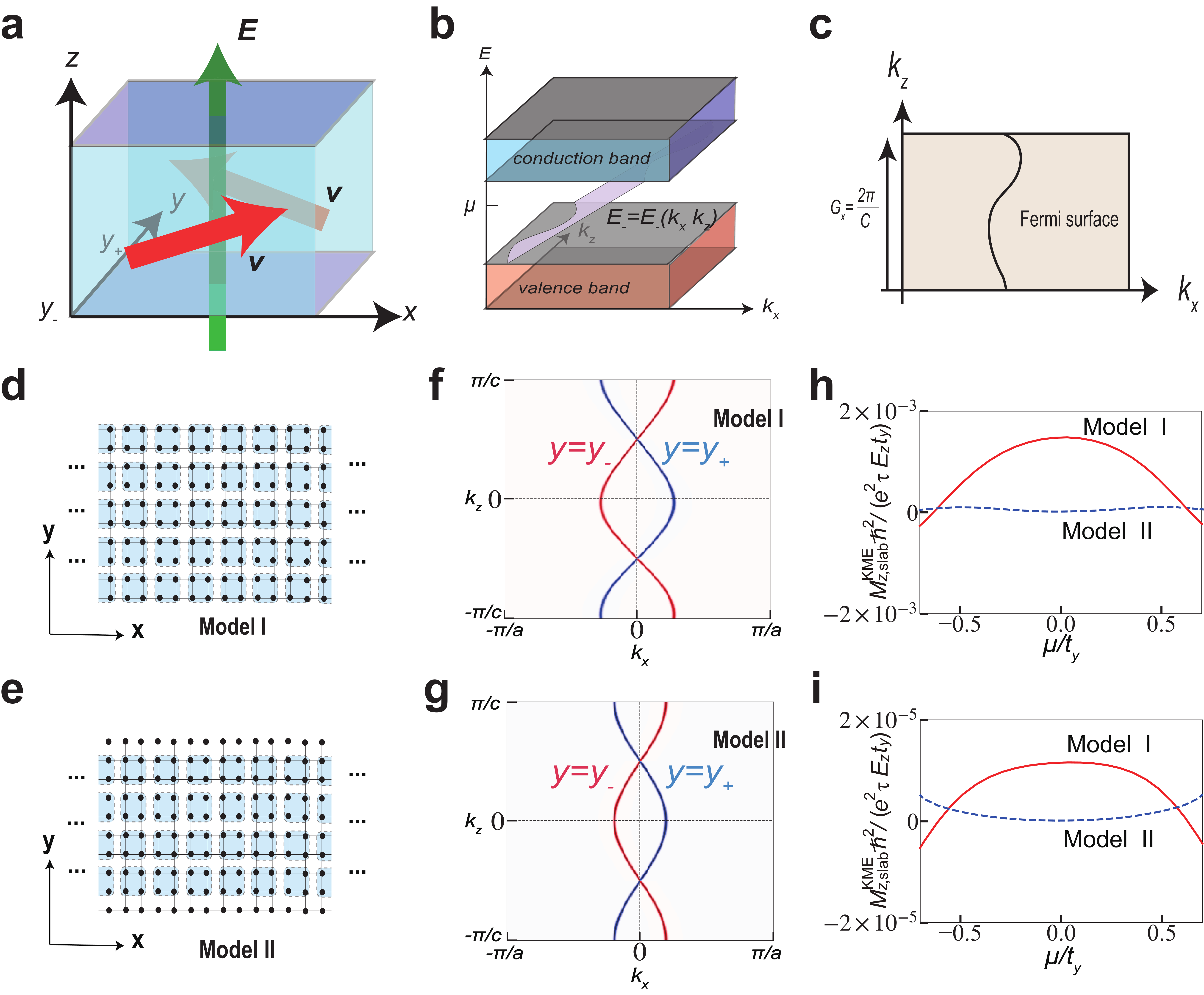}}
\caption{{\bf Chern insulator with a chiral structure and its kinetic magnetoelectric effect (KME) in slab systems.} ({\bf a}) Surface velocities (red arrow) of the topological surface states under an electric field ${\bf E}$(green arrow).  ({\bf b, c}) Band structure of the topological chiral surface states on the $y=y_-$ surface. Their ({\bf b}) dispersion and ({\bf c}) Fermi surface are shown.  In (b), the valence and conduction bands are represented by orange and blue boxes, and the topological surface states is drawn in purple in between the two bands. (c) is the Fermi surface, which is a section of (b) at the energy $E$ equal to the chemical potential $\mu$. 
({\bf d}) Slab models I and with chiral hopping on the surface. Only the section along the $xy$ plane is shown. The dots are the sites of the tight-binding model, and the blue boxes represent the unit cells, as shown in Fig.~1({\bf a}).
({\bf e})  Slab model I\hspace{-.1em}I with no chiral hopping on the surface. The symbols are the same as ({\bf d}). 
({\bf f, g}) The Fermi surfaces for the slab models ({\bf f}) I and ({\bf g}) I\hspace{-.1em}I with parameters $t_x=t_y=m=b_x=b_y$, $t_3=0.1t_y$, $t_4=0.15t_y$ and $\mu=0$.  The red and blue curves represent Fermi surfaces on the $y=y_-$ and $y=y_+$ planes, 
respectively. ({\bf h, i}) KME for the slab models I and I\hspace{-.1em}I, shown as red solid lines and as blue broken lines, respectively, with parameters ({\bf h}) $t_x=t_y=m=b_x=b_y$, $t_3=0.1t_y$ and $t_4=0.15t_y$ and ({\bf i}) $t_x=t_y=m=b_x=b_y$, $t_3=0.001t_y$ and $t_4=0.15t_y$.}
\label{f3}
\end{figure*}

\noindent{\bf Model calculation on a Chern insulator.}
As an example of a topological insulator, we consider an orbital magnetization in a Chern insulator with a chiral crystal structure. For this purpose, we introduce a three-dimensional tight-binding model of a layered Chern insulator, as shown in Fig.~\ref{f1}a, connected via right-handed interlayer chiral hoppings (Fig.~\ref{f1}b). Each layer forms a square lattice within the $xy$-plane, with a lattice constant $a$, and they are stacked along the $z$-axis with a spacing $c$. 
 as shown in detail in Methods.
The Brillouin zone and the band structure is shown in Figs.~\ref{f1}c-e.
We set the Fermi energy in the energy gap.

We calculate KME in a one-dimensional quadrangular prism with $xz$ and $yz$ surfaces shown in Fig.~\ref{f1}f  (see Methods),
with its results in Figs.~\ref{f1}g and \ref{f1}h with the interlayer hopping $t_3=0.1t_y$ and $t_3=0.001t_y$, respectively, for several values of the system size, $L_x$ and $L_y$, representing the lengths of the crystal in the $x$ and $y$ directions. Thus, the KME is affected by boundaries and system size, and this size dependence remains even when the system size is much larger than the penetration depth of topological surface states. Therefore, KME cannot be defined as a bulk quantity. 
{In contrast, the orbital magnetization in equilibrium is shown to be a
bulk quantity in crystals \cite{thonhauser,ceresoli,Xiao2005} in 2005, which is nontrivial, because the
operator for the orbital magnetization involves the position operator $\bf{r}$.} 
Later, we give an interpretation on this characteristic size dependence. 

\noindent{\bf Surface theory of KME for a slab.}
In topological insulators such as Chern insulators, only the topological surface states can carry a current. Here we calculate the KME using an effective Hamiltonian for the crystal surface. Thereby, we can capture natures of KME through this surface theory.
We consider slab systems, with its surfaces on $y=y_{\pm}$ $(y_+>y_-)$. The slab is sufficiently long along the $x$ and $z$ directions and we impose periodic boundary conditions in these directions. To induce the orbital magnetization $M_{z, {\rm slab}}^{\rm KME}$, we apply an electric field $E_z$  in the $z$ direction. Due to the interlayer chiral hoppings, the surface current acquires a nonzero $z$-component (Fig.~\ref{f3}a).

 Let $E_-=E_-(k_x, k_z)(=E_-(k_x, k_z+\frac{\pi}{c}))$ be the surface-state dispersion on the $y=y_-$ surface as shown in Fig.~\ref{f3}b and Fig.~\ref{f3}c. For simplicity, we assume $C_{2z}$ symmetry of the system. Then the surface state dispersion on the $y=y_+$ surface is given by
$E_+=E_+(k_x, k_z)=E_-(-k_x, k_z)$.
Here, we assume that the surface states are sharply localized at $y=y_{\pm}$, namely, we ignore finite-size effects due to a finite penetration depth. Then, we rewrite equation (\ref{3}) to
\begin{eqnarray}
&&M_{z,{\rm slab}}^{\rm KME}=\frac{e^2\tau E_z}{2\hbar^2}\int^{\pi/c}_{-\pi/c}\frac{dk_z}{(2\pi)^2}\nonumber\\
&&\ \ \ \times\frac{\partial E(k_x', k_z)}{\partial k_z}{\rm sgn}\left(\frac{\partial E(k_x', k_z)}{\partial k_x'}\right)\Big|_{E(k_x', k_z)=\mu},\label{Mslab3}
\end{eqnarray}
(see Supplementary Note 1 for details).
We note that the Fermi surface depends on the surface termination, and so does the KME. 
We also confirm the surface dependence from numerical calculations as shown in Figs.~\ref{f3}d-i.
This formula applies to any topological insulators such as $Z_2$-TIs\cite{Fu2007, Moore2007, Hsieh2008} (see Supplementary Note 2).

\begin{figure}
\raisebox{-\height}{\includegraphics[width=0.45\textwidth]{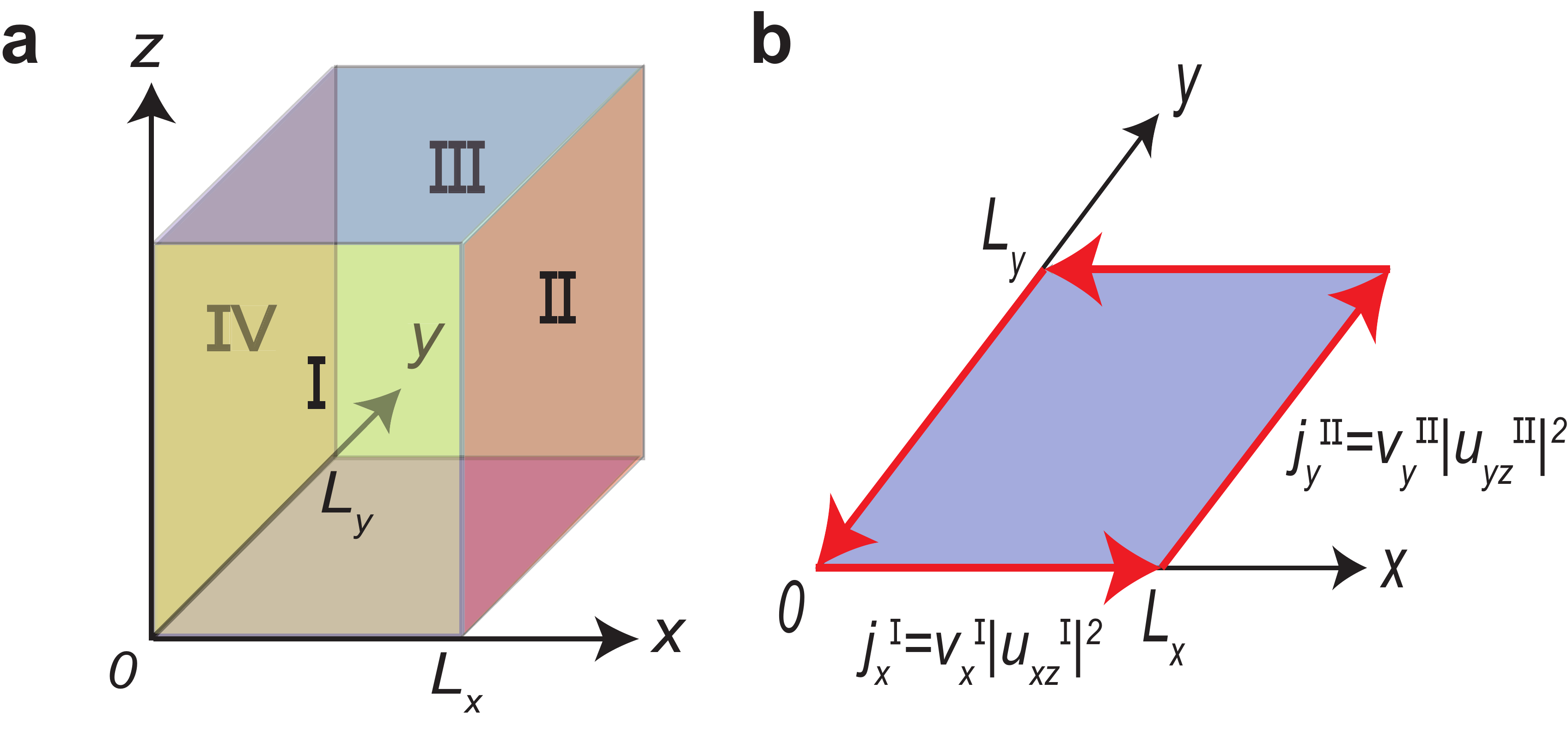}}
\caption{ {\bf One-dimensional prism of a Chern insulator. }({\bf a}) Schematic figure of the one-dimensional prism of a Chern insulator, extended along $z$-axis, with its size $L_x\times L_y$ within the $xy$ plane. We call the four side surfaces I, II, III and IV.  ({\bf b}) Schematic figure for current conservation at the corners around the one-dimensional prism in ({\bf a}). Here, $j_x^{{\rm I}}$ and 
$j_y^{{\rm II}}$ are current densities in the circumferential direction of the prism on the surfaces I and I\hspace{-.1em}I, respectively, and we impose them to be equal. 
$v_x^{{\rm I}}$ and 
$v_y^{{\rm II}}$ are corresponding velocities of electrons, and $u_{xz}^{{\rm I}}$ and 
$u_{yz}^{{\rm II}}$ are amplitudes of the wavefunctions in the respective surfaces.}
\label{f5}
\end{figure}

\noindent{\bf Surface theory of KME for a cylinder.}
From this slab calculation, we calculate the KME for a
cylinder geometry.
We consider a current along the $z$ direction in a one-dimensional quadrangular prism with $xz$ and $yz$ surfaces (surfaces I-I\hspace{-.1em}V in Fig.~\ref{f5}a) through its surface Hamiltonian.  Let $L_x$ and $L_y$ denote the system sizes
along the $x$ and $y$ directions, respectively.
Because the KME is sensitive to differences in crystal surfaces,
as shown in slab systems, we consider the individual surfaces separately.

In particular, in Chern insulators we can calculate the energy eigenstates for the whole system from those for the surface Hamiltonians. 
For simplicity, we assume twofold rotation symmetry $C_{2z}$ of the system, which relates between I and I\hspace{-.1em}I\hspace{-.1em}I, and between I\hspace{-.1em}I and I\hspace{-.1em}V. Then, only the surface I and I\hspace{-.1em}I are independent. We write down the eigenequations for these surfaces as
\begin{eqnarray}
H^{\rm I}\psi_{k_xk_z}^{\rm I}(x,z)&=&E^{\rm I}_{k_x k_z}\psi_{k_xk_z}^{\rm I}(x,z),\\
H^{\rm I\hspace{-.1em}I}\psi_{k_yk_z}^{\rm I\hspace{-.1em}I}(y,z)&=&E^{\rm I\hspace{-.1em}I}_{k_y k_z}\psi^{\rm I\hspace{-.1em}I}_{k_y k_z}(y,z),
\end{eqnarray}
where $H^{\rm I}$ and $H^{\rm I\hspace{-.1em}I}$ are the surface Hamiltonians for the surfaces I and I\hspace{-.1em}I, respectively, and 
\begin{math}
\psi_{k_xk_z}^{\rm I}=u_{k_xk_z}^{\rm I}(k_x,k_z)e^{ik_xx}e^{ik_zz}
\end{math}
and
\begin{math}
\psi_{k_yk_z}^{\rm I\hspace{-.1em}I}=u_{k_yk_z}^{\rm I\hspace{-.1em}I}(k_y,k_z)e^{ik_yy}e^{ik_zz}
\end{math}
are Bloch eigenstates on the surfaces I and I\hspace{-.1em}I, respectively. 
We can determine these eigenstates from four conditions, equality of the energy eigenvalues, current conservation at the corner (Fig.~\ref{f5}b) \cite{Raoux2010, concha, takahashi}, periodic boundary condition on the crystal surface and the normalization condition  (see Supplementary Note 3).

Thus we obtain a formula for KME in a one-dimensional prism of a three-dimensional Chern insulator
\begin{equation}
M_z^{\rm KME}=-\frac{e^2\tau E_z}{\hbar}\frac{1}{(2\pi)^2}\int^{\pi/c}_{-\pi/c}dk_z\frac{L_x\frac{\partial k_x^{{\rm I}}}{\partial k_z}+L_y\frac{\partial k_y^{{\rm I\hspace{-.1em}I}}}{\partial k_z}}{\frac{L_x}{v_x^{\rm I}}+\frac{L_y}{v_y^{\rm I\hspace{-.1em}I}}}\Big|_{E=\mu},
\label{Mone2}
\end{equation}
where $v_x^{\rm I}=\frac{1}{\hbar}\frac{\partial E^{\rm I}}{\partial k_{x}}$ and $v_y^{\rm I\hspace{-.1em}I}=\frac{1}{\hbar}\frac{\partial E^{\rm I\hspace{-.1em}I}}{\partial k_{y}}$ and
\begin{math}k_x^{{\rm I}}(k_z, E)
\end{math}
and 
\begin{math}
k_y^{{\rm I\hspace{-.1em}I}}(k_z, E)
\end{math}
are functions obtained from $E=E^{\rm I}_{k_xk_z}$ and $E=E^{\rm I\hspace{-.1em}I}_{k_yk_z}$, respectively.
When $v_x$ and $v_y$ are almost independent of $k_z$, we  approximate equation (\ref{Mone2}):
\begin{equation}
M_z^{\rm KME}=2\frac{\displaystyle\frac{L_x}{\langle v_x^{\rm I}\rangle}M^{\rm I, KME}_{z,{\rm slab}}+\frac{L_y}{\langle v_y^{\rm I\hspace{-.1em}I}\rangle}M^{\rm I\hspace{-.1em}I, KME}_{z,{\rm slab}}}{\displaystyle\frac{L_x}{\langle v_x^{\rm I}\rangle}+\frac{L_y}{\langle v_y^{\rm I\hspace{-.1em}I}\rangle}},
\label{Mone}
\end{equation}
where $M^{\rm I, KME}_{z,{\rm slab}}$ and $M^{\rm I\hspace{-.1em}I, KME}_{z,{\rm slab}}$ represent the KME for a slab (equation (\ref{Mslab3})) with the surface I and that with the surface I\hspace{-.1em}I, respectively. Thus, the KME of the one-dimensional system can be well approximated by equation (\ref{Mone}) expressed in terms of that for the slabs along $xz$ and along $yz$ planes.


In general topological insulators, we can also derive KME in terms of a simple picture of a combined circuit, consisting of four surfaces I- I\hspace{-.1em}V with anisotropic transport coefficients.
We obtain
\begin{equation}
M_z^{\rm KME}=j_{\rm circ}=\frac{\frac{L_x}{\sigma_{xx}^{\rm I}}\sigma_{xz}^{\rm I}+\frac{L_y}{\sigma_{yy}^{\rm I\hspace{-.1em}I}}\sigma_{yz}^{\rm I\hspace{-.1em}I}}{\frac{L_x}{\sigma_{xx}^{\rm I}}+\frac{L_y}{\sigma_{yy}^{\rm I\hspace{-.1em}I}}}E_z,
\label{Mone3}
\end{equation}
where $j_{\rm circ}$ is the circulating current density within the $xy$ plane around the prism per unit length along the $z$-direction. $\mathbf{\sigma}^{\rm I,I\hspace{-.1em}I}_{ij}$ is the electric conductivity tensor for the surfaces I and I\hspace{-.1em}I (see Supplementary Note 4).
On the other hand, we can also show
\begin{math}
M^{\rm I, KME}_{z,{\rm slab}}=\frac{1}{2}\sigma_{xz}^{\rm I}E_z, M^{\rm I\hspace{-.1em}I, KME}_{z,{\rm slab}}=\frac{1}{2}\sigma_{yz}^{\rm I\hspace{-.1em}I}E_z.
\end{math}
In Chern insulators, by using
\begin{math}
\sigma_{xx}^{\rm I}\propto \langle v_x\rangle
\end{math}
and
\begin{math}
\sigma_{yy}^{\rm I\hspace{-.1em}I}\propto \langle v_y\rangle,
\end{math}
we arrive at equation (\ref{Mone}). Thus, we can calculate the KME from the surface electrical conductivity from equation (\ref{Mone3}), which depends on the aspect ratio $L_x/L_y$.

\begin{figure}
\raisebox{-\height}{\includegraphics[width=0.5\textwidth]{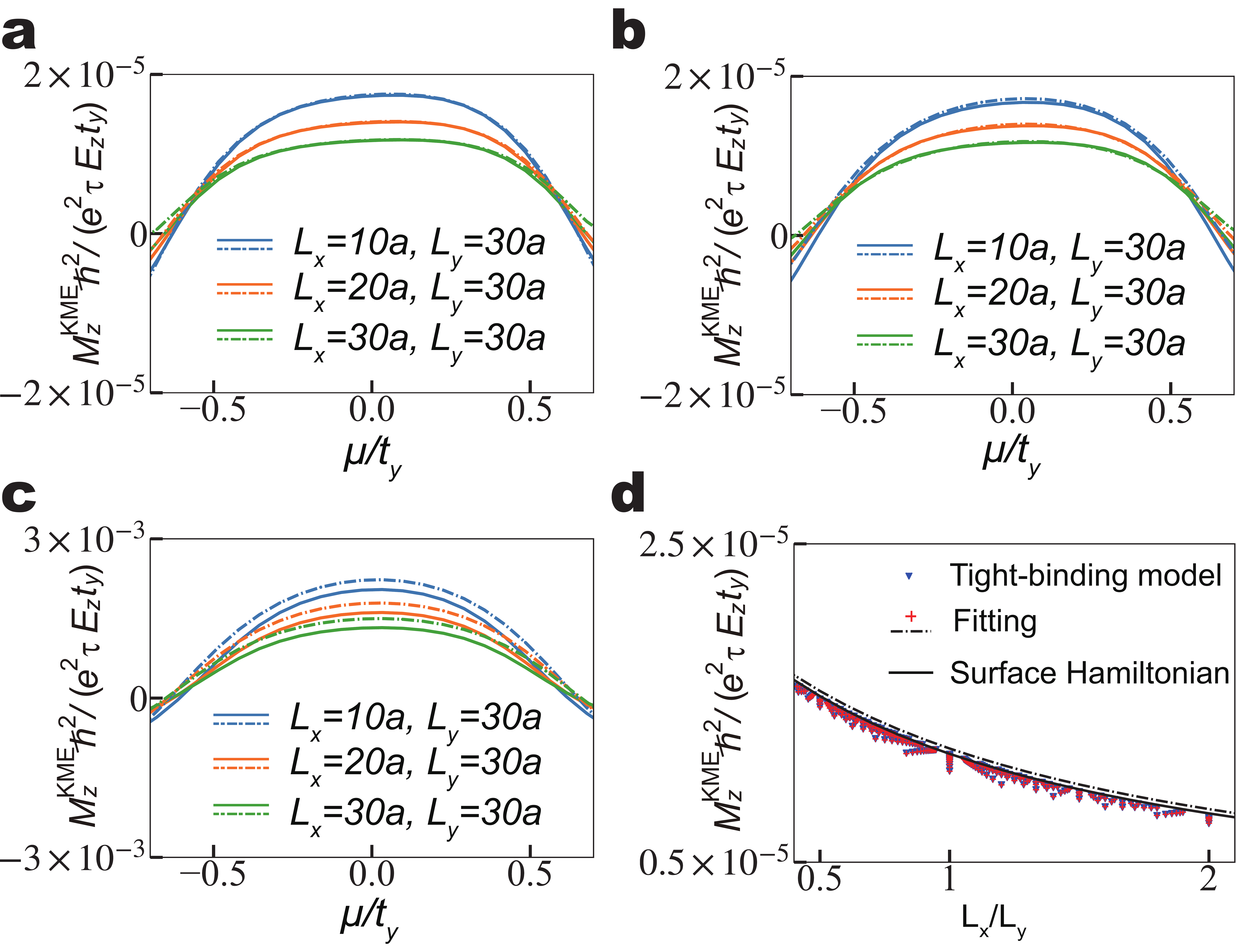}}
\caption{ {\bf Kinetic magnetoelectric effect (KME) in one-dimensional systems. }({\bf a-c}) KME calculated from two different methods; one is a direct calculation by equation (\ref{3}) (solid lines) and the other is by a combination of calculation results for surfaces along $xz$ and $yz$ planes based on equation ({\ref{Mone}}) (dashed lines). Parameter values are ({\bf a}) $t_x=t_y=m=b_x=b_y$, $t_3=0.001t_y$ and $t_4=0.15t_y$, ({\bf b}) $t_x=b_x=1.1t_y$, $t_y=b_y=m$, $t_4=0.15t_y$ and $t_3=0.001t_y$ and ({\bf c}) $t_x=t_y=m=b_x=b_y$, $t_4=0.15t_y$ and $t_3=0.1t_y$. The results are shown for $L_x=10a$ (blue), $L_x=20a$ (orange) and $L_x=30a$ (green), while $L_y$ is fixed to be $30a$. 
({\bf d}) Dependence of the KME on the aspect ratio within the $xy$ plane with parameters $t_x=b_x=1.1t_y$, $t_y=b_y=m$, $t_4=0.15t_y$, $t_3=0.001t_y$ and $\mu=0$. Blue points represent the result of direct calculation by equation (\ref{3}) for various system sizes. With parameter values $t_x=b_x=1.1t_y$, $t_y=b_y=m$, $t_4=0.15t_y$, $t_3=0.001t_y$ and $\mu=0$, we calculate the KME with various system sizes from equation (\ref{3}). The system sizes are $(L_x, L_y)\in \{10a,12a,\dots,38a\}\times\{10a,12a,\dots,38a\}$. Red points represent the result of fitting the numerical results of equation (\ref{3}) with the fitting function in equation (\ref{mk}), and its limit for $L_x$, $L_y$$\rightarrow \infty$ is shown as the dashed line. The solid line represents the results of the surface theory in equation (\ref{Mone}).}
\label{f7}
\end{figure}

We numerically comfirm that the results of direct calculation by equation (\ref{3}) and those for  surface calculation by equation (\ref{Mone}) agree well (Fig.~\ref{f7}a-c). When the interlayer hopping is large (Fig.~\ref{f7}c), they slightly deviate from each other. This is because we cannot ignore the $k_z$ dependence of $v_x$ and $v_y$ and they are out of the scope of the approximate expression (\ref{Mone}).

\noindent{\bf Finite-size effect.} In our approximation theory, we assumed that the surface current is localized at the outermost sites and ignored a finite penetration depth. In fact, we can fit well the data with various system sizes with a trial fitting function which includes a finite-size effect in equation (\ref{Mone}) (see Methods and Supplementary Note 5). From these results, the finite-size effect is of the order $1/L$ in the leading order, coming from the finite penetration depth. When the system size is much larger than the penetration depth, the result is well described by the surface theory as shown in Fig.~\ref{f7}d.

\noindent{\bf Materials.} 
Topological insulators without inversion symmetry can be a good platform for obtaining large KME, because the current flows on the surface. Therefore, the closed loop created by the current is macroscopic and it efficiently induces the orbital magnetization. 
In contrast, in the conventional KME in metals, a bulk current generates microscopic current loops in the bulk, which leads to a much smaller effect than a surface current in topological insulators. One can also regard  
this set of current loops as a macroscopic current loop along the surface, but the current in this case is of a microscopic amount, determined by the 
current per bulk unit cell. Therefore, the resulting effect in bulk metals is much smaller than the KME in topological insulators, where the current along the surface is of a macroscopic size. Moreover, the surface states of topological materials are robust against perturbations caused by impurities. 

\begin{figure*}
\raisebox{-\height}{\includegraphics[width=1\textwidth]{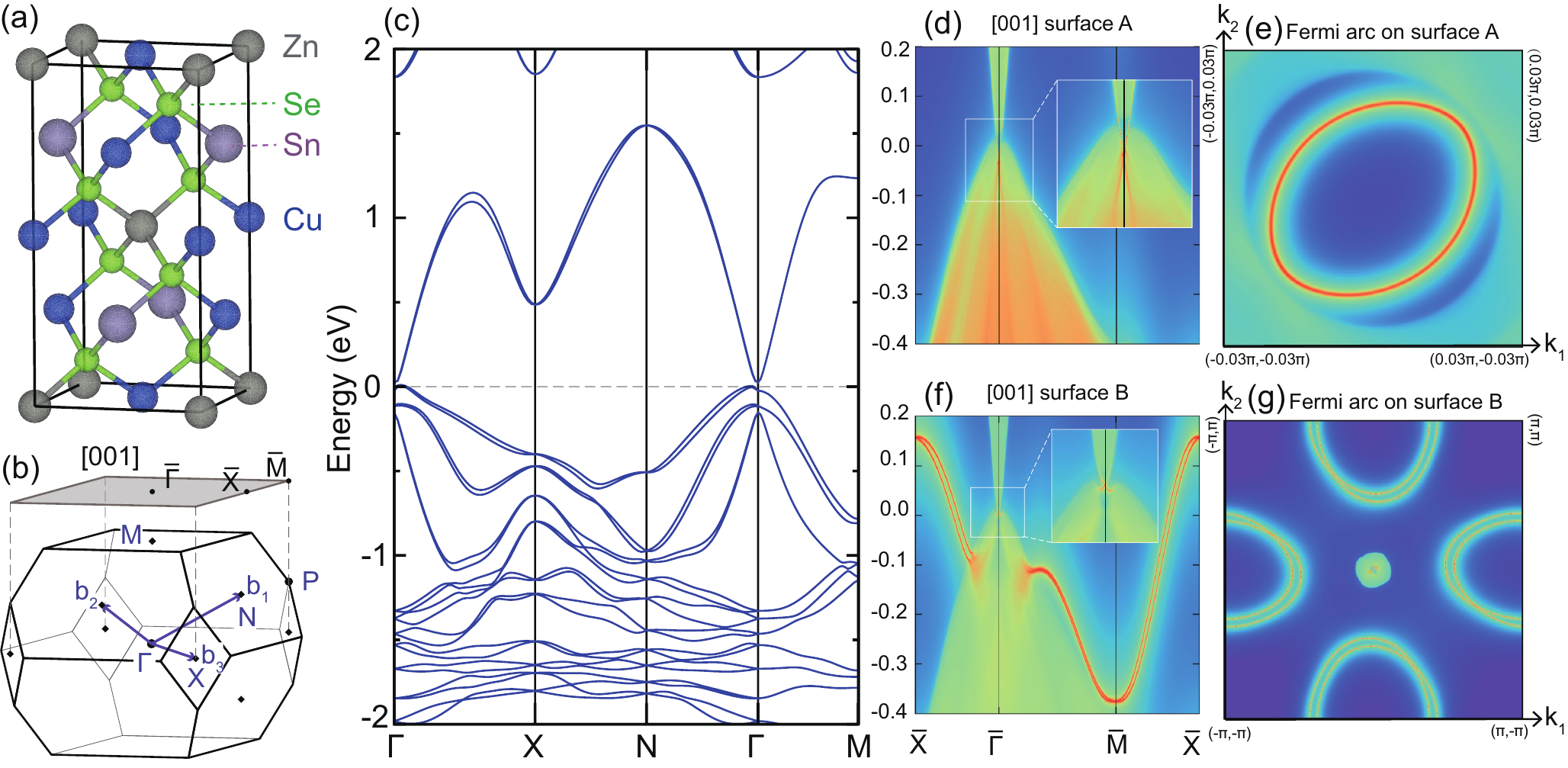}}
\caption{{\bf First-principle calculations on Cu$_2$ZnSnSe$_4$.} ({\bf a}) Crystal structure of Cu$_2$ZnSnSe$_4$. ({\bf b})  Brillouin zone and surface Brillouin zone along [001] direction. ({\bf c}) Electronic structure with spin-orbit coupling for the bulk. ({\bf d-e}) Surface states and Fermi arcs calculation on the [001] surface with Cu-Sn layer termination (surface A). ({\bf f-g}) Surface states and Fermi arcs calculation on the [001] surface with Se layer termination  (surface B). }
\label{f4}
\end{figure*}

Under the non-inversion-symmetry constraint, 
we cannot diagnose $Z_2$-TIs easily because the $Z_2$ topological invariant
is expressed in terms of $k$-space integrals.
Our idea here is to use S$_{4}$ symmetry to diagnose $Z_2$-TIs,
where we only need to calculate wavefunctions at four momenta according to the symmetry-based indicator theories \cite{po2017symmetry,song2018quantitative,khalaf2018symmetry}. After searching in the topological material database \cite{zhang2019catalogue}, we notice that Cu$_2$ZnSnSe$_4$ \cite{guen1979physical} with $\bm{{\it 82}}$ and CdGeAs$_{2}$ with $\bm{{\it 122}}$ are two ideal candidates of $Z_2$-TIs with a direct gap for obtaining a large KME (see Supplementary Note 6 for details).
In the following, we will use Cu$_2$ZnSnSe$_4$, which only has $S_{4}$ symmetry as shown in Fig.~\ref{f4}a, as an example to show the magnitude of the KME with different surfaces and different surface terminations.

Since the magnetoelectric tensor for the space group $\bm{{\it 82}}$, defined by 
$\bm{M}=\bm{\alpha}\bm{E}$, is $\alpha_{\bm{{\it 82}}}= \left[
 \begin{matrix}
   \alpha_{11} & \alpha_{12} & 0 \\
   \alpha_{12} & -\alpha_{11} & 0 \\
   0 & 0 & 0
  \end{matrix}  \right] $, we can obtain an orbital magnetization $M^{\rm KME}_{1}$ by adding an external electric field $E_{1}$, through 
  the surface currents both on the [001] surface and on the [010] surface thanks to the nonzero $\alpha_{11}$ (see Supplementary Note 7 for details). In our discussion of the KME effect, we set the current direction to be along $z$ axis. Therefore we
  will set the 1-axis in the above magnetoelectric tensor to be the $z$ axis in our theory.

Figures~\ref{f4}b and c are the Brillouin zone and the band structure of Cu$_2$ZnSnSe$_4$ with a gap,
through the first-principle calculations whose  details are
explained in Methods. On the [001] surface, terminations with Cu-Sn layer (surface A) and with Se layer (surface B) have different surface energies and Fermi surfaces, as shown in Fig.~\ref{f4}d-g, which contribute to a magnetoelectric susceptibility of  $\alpha_{11}^{\rm A}=-1.804\times10^{8} s^{-1}\Omega^{-1}$ $\cdot \tau$ and $\alpha_{11}^{\rm B}=-2.565\times10^{9} s^{-1}\Omega^{-1}$ $\cdot \tau$, respectively. 
On the A surface, there is a single surface Dirac cone at $\Gamma$ point, forming an electron-like Fermi surface. On the B surface, the Dirac cone at $\Gamma$ point forms an almost zero Fermi surface, but two surface Dirac cones at two $\bar{X}$ momenta form two hole-like Femi surfaces. 
Because the Fermi surfaces on the B surface are much larger than those on the A surface, the magnetoelectric susceptibility on the B surface is one order of magnitude larger than that on the A surface. 
Similar calculations on the [010] surface are in the Supplementary Note 8, and the
result is $\alpha_{11}^{\rm C}=-2.324\times 10^{9} s^{-1}\Omega^{-1}$ $\cdot \tau$
for the surface C.

Let us compare the results with metallic materials in the bulk. 
For simplicity, we focus on the cases with the electric field $\bm{E}$ and 
the resulting magnetization $\bm{M}^{\rm KME}$ along the $z$ direction. 
This kinetic magnetoeletric response is expressed as
$M_z^{\rm KME}=\alpha_{zz} E_z$, and the conductivity is 
$j_z=\sigma_{zz} E_z$. Thus the magnetization in response to the current is
$M_z^{\rm KME}=(\alpha_{zz}/\sigma_{zz})j_z$. In the relaxation time approximation, both 
$\alpha_{zz}$ and $\sigma_{zz}$ are proportional to the relaxatoin time $\tau$.
For simplicity we consider the system to be a cube with 
its size $L\times L\times L$. 
In the bulk metallic systems, the current is carried by the bulk states, 
and $\alpha_{zz}\propto L^0$, $\sigma_{zz}\propto L^0$. 
On the other hand, 
in topological systems, only the surface conducts the current, 
and the conductivity $\sigma_{zz}$ scales as  $\sigma_{zz}\propto L^{-1}$. 
On the other hand, we have shown $\alpha_{zz}\propto L^0$, which means that $\alpha_{zz}$ is an intensive quantity. 
Thus, in topological insulators, the scaling of the KME as a response to the electric field is represented by the response coefficient $\alpha_{zz}\propto L^{0}$.
Meanwhile, the response coefficient $\alpha_{zz}/\sigma_{zz}$
to the current is proportional to $L$. It means that as
a response to the current, topological materials will generate a large 
amount of orbital magnetization as compared to metals.

We compare our results with KME
in p-doped tellurium, which has chiral crystal structure \cite{Tsirkin2018}. 
For the acceptor concentration $N_a=4\cdot 10^{14}{\rm cm}^{-3}$ 
at 50K, the induced orbital magnetizations is
$M^{\rm KME}_z=7.0\cdot 10^{-8}$ $\mu_{\rm B}/{\rm atom}\sim 1.85\times10^{-2}$A/m by a current density $j_z=1000$A/cm$^2$. Thus the response coefficient of the orbital magnetization $M^{\rm KME}_z$ to
the current density $j_z$ is $\alpha_{zz}/\sigma_{zz}=1.85\times 10^{-9}$m. 
If we approxima1te $\sigma_{zz}$ by $\sigma_{zz}\sim \frac{N_ae^2\tau}{m}$ with
the electronic charge $e$ and mass $m$, we get
$\alpha=2.1\times 10^4$ s$^{-1}\Omega^{-1}$ $\cdot \tau$.
For other acceptor concentrations
$N_a=4\cdot 10^{16}{\rm cm}^{-3}$  and
$N_a=1\cdot 10^{18}{\rm cm}^{-3}$, one can similarly 
get $\alpha_{zz}=2.1\times 10^6$ s$^{-1}\Omega^{-1}$ $\cdot \tau$ and 
$\alpha_{zz}=2.3\times 10^7$ s$^{-1}\Omega^{-1}$ $\cdot \tau$.
Thus, the size of $\alpha_{zz}$ for the topological insulator Cu$_2$ZnSnSe$_4$ 
is larger than that of Te by two to five orders of magnitude.

On the other hand, in topological insulators, the induced orbital magnetization as a response to the current becomes huge compared with metals.
To show this, we consider a system with surfaces having anisotropic transport coefficients with 
sheet resistance
 $\sigma_{xz,\square}$ and $\sigma_{zz,\square}$. Then we can define an angle $\theta$ by
 $\sigma_{xz,\square}/\sigma_{zz,\square}=\tan\theta$, where $\theta$ describes an angle between the electric field along the $z$ direction and the surface current density $\bm{j}^{\text{surf}}$.
For example, for the [001] surface of Cu$_2$ZnSnSe$_4$,
we get $\sigma_{21,\square}^{\rm A}=2\alpha_{11}^{\rm A}=-3.6\times10^{8} s^{-1}\Omega^{-1}$ $\cdot \tau$, $\sigma_{11,\square}^{\rm A}=7.3\times10^{8} s^{-1}\Omega^{-1}$ $\cdot \tau$, $\sigma_{21,\square}^{\rm B}=2\alpha_{11}^{\rm B}=-5.1\times10^{9} s^{-1}\Omega^{-1}$ $\cdot \tau$, $\sigma_{11,\square}^{\rm B}=2.0\times10^{10} s^{-1}\Omega^{-1}$ $\cdot \tau$,
which yield $\tan\theta^{\rm A}=-0.49$ and $\tan\theta^{\rm B}=-0.26$ by identifying $x=2$ and $z=1$.
Then the total current along the $z$ direction is $4Lj_z$ while the circulating current is $j_{\rm circ}=j^{\text{surf}}_x=j^{\text{surf}}_z\tan \theta$. Thus the magnetization response $M^{\rm KME}_z$ to the current density $j_z(=4Lj_z^{\text{surf}}/L^2)$ is $M_z^{\rm KME}/j=(j_{\rm circ}/j^{\text{surf}}_z)(L/4)=(L/4)\tan\theta$, which is proportional to the system size. {
It is the response coefficient $\alpha/\sigma$ shown in Table I. }
Thus for the macroscopic system size, 
the response $M^{\rm KME}_z/j_z$ is also of the macroscopic size such as milimeters,
and it is many order of magnitude larger than that in tellurium, where $M_z^{\rm KME}/j_z$ is evaluated to be $M_z^{\rm KME}/j_z=1.85\times 10^{-9}$m.

{
This scaling to the system size shows a prominent difference in the KME in topological insulators from similar effects.
In bulk metals studied in previous works, the KME is always independent of the system size $L$. In topological insulators, the
current induces both the spin and the orbital magnetizations, and we propose that only the orbital magnetization shows a different scaling behavior. As a comparison, we calculate the spin magnetization induced by the 
electric field in Cu$_2$ZnSnSe$_4$, with calculation details and spin textures presented in Supplementary Note 9, and obtain $\alpha_{11}^{\text{A,spin}}=2.798\times10^{-2} \text{m}\cdot \text{s}^{-1}\Omega^{-1}$ $\cdot \tau/L_y$, and
$\alpha_{11}^{\text{B,spin}}=-2.575 \text{m}\cdot \text{s}^{-1}\Omega^{-1}$ $\cdot \tau/L_y$.
Thus it inversely scales with the system size along the $y$ direction.
Thus suppose the system size of 1mm, they are of the order $10^1-10^3\text{s}^{-1}\Omega^{-1}$ $\cdot \tau$, and the orbital magnetization from the KME, 
having $10^8-10^9\text{s}^{-1}\Omega^{-1}$ $\cdot \tau$,
is larger than the spin counterpart by 5 - 8 orders of magnitude.
Thus, while the spin and orbital magnetization behave similarly in bulk metals, 
they are quite different in topological insulators, which is the main point of the present paper. }

\noindent{\bf Conclusion}\\
In summary, we propose KME in topological insulators with chiral structure. This KME is carried by surface current due to the asymmetric crystal structure of the surface. Therefore, the KME is sensitive to surface terminations, and it cannot be defined as a bulk quantity. We derive a formula for the KME as a surface quantity using the surface Hamiltonian, and show that it fits with numerical results.

In theoretical treatments, atomic orbitals can classify 
the orbital magnetization into intraatom and interatom
contributions. 
Some atomic orbitals such as $p_x\pm ip_y$ have orbital angular momentum, which leads to corresponding intraatomic orbital magnetization.
On the other hand, the hopping between atoms lead, to the interatomic 
orbital magnetization.
In tight-binding models with atomic orbitals, they are separately calculated.
In some papers \cite{Shalygin2012, Koretsune2012}, the intraatomic orbital magnetization is studied, while 
the interatomic one is studied in other papers \cite{yoda1,yoda2, Hara2020}
In real mateirals, these two contributions are not separable, and 
in the {\it ab initio} calculation \cite{Tsirkin2018}, their sum is calculated. 
In this paper, we found that in topological materials, the interatomic contribution is much larger due to the macroscopic current loop.
We show that the response to the current in topological insulators is much larger than in metals.
{In Table~\ref{tab:scaling}, we show scaling behaviors of
the spin and intraatom/interatom orbital magnetizations for metals and TIs. The KME is shown as
responses to an electric field, $\alpha$, and to a current, $\alpha/\sigma$. In particular, in the 
response to a current, $\alpha/\sigma$, it scales as $L^1$ only in the interatom orbital magnetization, while other entities scales with $L^0$. This shows a particular feature of the interatom orbital magnetization generated by a current proposed in the present paper.}

{In this paper, we put two $Z_2$ TIs as candidates for the topological KME.
In addition to Chern insulators and $Z_2$ TIs, other classes of topological insulators such as
various classes of topological crystalline insulators, and topolgical semimetals will also show the topological KME. In topological semimetals, topological surface states coexist with bulk metallic states, but the contribution from the former overwhelms that from the latter, and the topological KME is expected.}

\noindent{\bf Methods}\\
\noindent{\bf Details of the first-principle calculations}
First-principle calculations of Cu$_2$ZnSnSe$_4$ are implemented in the Vienna $ab$ $initio$ simulation package (VASP) \cite{Kresse1993,Kresse1994,Kresse1996} with  Perdew-Burke-Ernzerhof exchange correlation. A $\Gamma$-centered Monkhorst-Pack grid with 10$\times$10$\times$10 $k$-points and 460.8 eV for the cut-off energy of the plane wave basis set is used for the self-consistent calculation. Surface states and Fermi surfaces calculations are performed by the tight-binding model obtained by the maximally localized Wannier functions \cite{mostofi2008wannier90}.  

\noindent{\bf Details of the model Hamiltonian.}
We consider a Chern insulating system with a chiral crystal structure. 
The model is composed of infinite layers of the two-dimensional Wilson-Dirac model \cite{creutz,yoshimura}.
The lattice sites are expressed by $(i, j, l)$, with $i$, $j$, $l$ being integers, specifying the $x$, $y$ and $z$-coordinates. At each lattice site, we consider two orbitals $1$ and $2$. Let $c_{i,j,l,\sigma}$ denote the annihilation operator of electrons at the $(i, j, l)$-site with orbital $\sigma(=1,2)$, and we write
\begin{math}
c_{i,j,l}=
\left(c_{i,j,l,1},c_{i,j,l,2}\right)^T
\end{math}.
The model Hamiltonian is 
\begin{math}
H=H_{\rm WD}(m, t_x, t_y, b_x, b_y)+H_{\rm interlayer}(t_3, t_4)
\end{math}
, where $H_{\rm WD}$ is an in-plane Wilson-Dirac Hamiltonian, and $H_{\rm interlayer}$ is an interlayer Hamiltonian representing a structure similar to right handed solenoids.
The in-plane Wilson-Dirac Hamiltonian is
\begin{eqnarray}
&&H_{\rm WD}
=m\sum_{i,j,l}c^{\dagger}_{i,j,l}\sigma_zc_{i,j,l}\nonumber\\
&&\ \ -\frac{it_x}{2}\sum_{i,j,l}(c^{\dagger}_{i,j,l}\sigma_xc_{i+1,j,l}-H.c.)\nonumber\\
&&\ \ -\frac{it_y}{2}\sum_{i,j,l}(c^{\dagger}_{i,j,l}\sigma_yc_{i,j+1,l}-H.c.)\nonumber\\
&&\ \ +\frac{b_x}{2}\sum_{i,j,l}(c^{\dagger}_{i,j,l}\sigma_zc_{i+1,j,l}+H.c.-2c^{\dagger}_{i,j,l}\sigma_z c_{i,j,l})\nonumber \\
&&\ \ +\frac{b_y}{2}\sum_{i,j,l}(c^{\dagger}_{i,j,l}\sigma_zc_{i,j+1,l}+H.c.-2c^{\dagger}_{i,j,l}\sigma_z c_{i,j,l}), 
\end{eqnarray}
where H.c. stands for Hermitian conjugate of the preceding terms, $\dagger$ represents Hermitian conjugate, and
\begin{math}
m,t_x,t_y,b_x
\end{math}
and $b_y$ are real parameters. This Hamiltonian $H_{\rm WD}$ can be rewritten in the momentum space as
\begin{eqnarray}
\tilde{H}_{\rm WD}({\bf k})&=&t_x\sin{k_xa}\sigma_x+t_y\sin{k_ya}\sigma_y\nonumber\\
&+&(m-b_x(1-\cos{k_xa})-b_y(1-\cos{k_ya}))\sigma_z,\nonumber\\
\end{eqnarray} 
where ${\bf k}$ is the Bloch wavenumber.
An isotropic version of the two-dimensional Wilson-Dirac model with $b\equiv b_x=b_y$ and $t_x=t_y$ exhibits the Chern insulating phase when
\begin{math}
0<m/b<2
\end{math}
and
\begin{math}
2<m/b<4
\end{math}
 \cite{yoshimura}.
Next we add interlayer hoppings, including a direct hopping $t_4$ along the $z$-axis and a chiral hopping $t_3$, where $t_3$ and $t_4$ are real parameters. To describe the chiral hopping $t_3$, the lattice sites in the square lattice in each layer into groups of four sites, 
\begin{math}
(2i-1,2j-1), (2i-1,2j), (2i, 2j-1)
\end{math}
and
\begin{math}
(2i,2j)
\end{math}
where $i$ and $j$ are integers, and we introduce chiral hoppings between the groups on the neighboring layers. Then the total Hamiltonian for this model on a tetragonal lattice is given by 
\begin{equation}
H=H_{\rm WD}+H_{\rm interlayer},
\label{H}
\end{equation}
where
\begin{eqnarray}
H_{\rm interlayer}
&=&t_{3}\sum_{i,j,l}(c^{\dagger}_{2i-1,2j-1,l}c_{2i,2j-1,l+1}+H.c.)\nonumber\\
&&+t_{3}\sum_{i,j,l}(c^{\dagger}_{2i,2j-1,l}c_{2i,2j,l+1}+H.c.)\nonumber\\
&&+t_{3}\sum_{i,j,l}(c^{\dagger}_{2i,2j,l}c_{2i-1,2j,l+1}+H.c.)\nonumber\\
&&+t_{3}\sum_{i,j,l}(c^{\dagger}_{2i-1,2j,l}c_{2i-1,2j-1,l+1}+H.c.)\nonumber\\
&&+t_4\sum_{i,j,l}(c^{\dagger}_{i,j,l}c_{i,j,l+1}+H.c.).
\end{eqnarray}
These hoppings in 
\begin{math}
H_{\rm interlayer}
\end{math}
form structures similar to right-handed solenoids. 
When $H_{\rm WD}$ in the Chern insulator phase, even if $H_{\rm WD}$ is perturbed by $H_{\rm interlayer}$, the system remains in the Chern insulator with the Chern number within the $xy$ plane equal to $-1$ as long as $t_3$ and $t_4$ are small. In the main text, we are interested in the KME due to the topological surface states in the topological Chern insulating phase, in which the Fermi energy is in the energy gap.

\noindent{\bf Fitting function for KME.}
By taking into account the finite-size effect in equation (\ref{Mone}), we give a fitting fuction.
The finite penetration depth of the surface states will lead to $O(1/L)$ correction to the KME, and that around the corner will lead to $O(1/L^2)$ correction \cite{ceresoli}.
Thus, the fitting function is
\begin{equation}
M_z^{\rm KME}=\frac{w_1L_x+w_2L_y+w_3+\frac{w_4}{L_x}+\frac{w_5}{L_y}}{w_6L_x+w_7L_y},
\label{mk}
\end{equation}
where $w_i (i=1,2,\dots,7)$ are real constants.

\vspace{2mm}

\noindent
{\bf Acknowledgement}\\
This work was supported by Japan Society for the Promotion of Science (JSPS) KAKENHI Grants No. JP18H03678, No. JP20H04633, and No. JP21K13865 and by Elements strategy Initiative to Form Core Research Center (TIES), from MEXT Grant Number JP-MXP0112101001.

\vspace{2mm}

\noindent
{\bf Additional information}\\
The authors declare no competing financial interests.

\vspace{2mm}

\noindent
{\bf Data availability statement}\\
The datasets generated during and/or analysed during the current study are available from the corresponding authors on reasonable request.

\vspace{2mm}

\noindent
{\bf Code availability statement}\\
The source code for the calculations performed in this work is available from the corresponding authors upon reasonable request.

\vspace{2mm}

\noindent
{\bf Author contribution}\\
All authors contributed to the main contents of this work.
K. O. performed the model calculation and formulated the theory for topological systems, through the discussions with T. Z. and S. M.
T. Z. performed  
the {\it ab initio} calculation.
S.M. conceived and supervised the project.  
All authors drafted the manuscript.

\vspace{2mm}

\noindent{\bf References}\\

\clearpage

\begin{table}
\begin{tabular}{c|c||ccc}
\multicolumn{2}{c||}{}& $\sigma$ &  $\alpha$ &$\alpha/\sigma$\\
\hline
\multirow{2}{*}{interatom orbital} &
metal& $L^0$ & $L^0$ & $L^0$\\
\cline{2-5}&TI& $L^{-1}$ & $L^0$ & $L^1$\\
\hline
\multirow{2}{*}{intraatom orbital} &
metal& $L^0$ & $L^0$ & $L^0$\\
\cline{2-5}&TI& $L^{-1}$ & $L^{-1}$ & $L^0$\\
\hline
\multirow{2}{*}{spin} &
metal& $L^0$ & $L^0$ & $L^0$\\
\cline{2-5}&TI& $L^{-1}$ & $L^{-1}$ & $L^0$\\
\hline
\end{tabular}
\caption{Scaling behaviors of inter- and intraatom orbital and spin magnetizations versus the system size $L$, as a response to the electric field represented by $\alpha$, and to the current $\alpha/\sigma$ in metals and in topological insulators (TIs). The scaling of the conductivity $\sigma$ is also shown. }
\label{tab:scaling}
\end{table}
\end{document}



\title{Kinetic magnetoelectric effect in topological insulators: Supplemental material}%

\author{Ken Osumi}%
\affiliation{Department of Physics, Tokyo Institute of Technology, Tokyo 152-8551, Japan}
\author{Tiantian Zhang}%
\affiliation{Department of Physics, Tokyo Institute of Technology, Tokyo 152-8551, Japan}
\affiliation{TIES, Tokyo Institute of Technology, Tokyo 152-8551, Japan}
\author{Shuichi Murakami}%
\email{murakami@stat.phys.titech.ac.jp}
\affiliation{Department of Physics, Tokyo Institute of Technology, Tokyo 152-8551, Japan}
\affiliation{TIES, Tokyo Institute of Technology, Tokyo 152-8551, Japan}
\date{\today}%


\maketitle


\noindent
{\bf Supplementary Note 1. Surface theory in a slab of a 3D Chern insulator}

In this section, for a slab of a 3D Chern insulator we explain how to rewrite Eq.~(2) into Eq.~(3) in the main text.
Let us set the Chern number along the $xy$ plane to be $C_{xy}=-1$. The energy of the chiral surface states on the two surfaces $y=y_{\pm}$ based only on an intralayer Hamiltonian within the $xy$ plane, representing the Chern insulator, is
\begin{math}
E_{\pm}=\mp \hbar v_xk_x
\end{math}
\begin{math}
(v>0),
\end{math}
where $k_x$ is the wavevector along the $x$-direction and $v_x$ is the velocity along the $x$-direction. It means that the surface current is along the $x$-direction. When we include the interlayer Hamiltonian having chiral hoppings, the surface current on the surface will acquire a nonzero $z$-component as shown in Fig.~2a. Therefore, the energies of the chiral surface states of this model on the surface $y=y_{\pm}$ are
\begin{equation}
E_{\pm}(k_x, k_z)=\mp \hbar v_xk_x+\hbar v_zk_z.
\label{E}
\end{equation}
Here, the velocity $v_z$ along the $z$-direction is added to represent chiral nature of the hopping along the surface. This simple form means that the velocity of the chiral surface states is a constant value $(v_x, v_z)$. Nonetheless, this formula needs to be modified by the following reason. The energy should be a periodic function of $k_z$ with a period of the reciprocal lattice vector $G_z=\frac{2\pi}{c}$ along the $z$-direction:
\begin{math}
E(k_x, k_z)=E(k_x, k_z+\frac{2\pi}{c}),
\label{vk}
\end{math}
but Eq.~(\ref{E}) does not satisfy this requirement and we cannot use Eq.~(\ref{E}). On the other hand, since $E(k_x, k_z)$ represents the topological chiral surface states, it is not periodic along the $k_x$ direction. For example, on the $y=y_-$ surface, by increasing $k_x$ the surface states come out of the bulk valence band, and finally go into the bulk conduction band, as shown in Fig.~2b. Let $E_-=E_-(k_x, k_z)$ be the surface-state dispersion on the $y=y_-$ surface. For simplicity, we assume $C_{2z}$ symmetry of the system. Then the surface state dispersion on the $y=y_+$ surface is given by
\begin{eqnarray}
E_+=E_+(k_x, k_z)=E_-(-k_x, k_z).
\end{eqnarray}

Let
\begin{math}
\psi_{+(-)}(k_x, k_z)
\end{math}
denote a wave function on the $y=y_+$ $(y=y_-)$ surface of the crystal. 
In this calculation, we assume that the surface states are sharply localized at $y=y_{\pm}$, which means that we ignore finite-size effects due to a finite penetration depth.
From Eq.~(3), we evaluate the KME as
\begin{eqnarray}
\displaystyle
M_{z, {\rm slab}}^{\rm KME}&=&M_{z, +}+M_{z, -},\\
M_{z, +}&=&\frac{1}{(2\pi)^2}\int^{\pi/c}_{-\pi/c}dk_z\frac{1}{L_y}\int^{\pi/a}_{-\pi/a}dk_x \frac{e^2\tau E_z}{2\hbar}\frac{\partial f^0(E_+(\mathbf{\tilde{k}}))}{\partial k_z}y_+v_{x, +}(\mathbf{\tilde{k}}),\\
M_{z, -}&=&\frac{1}{(2\pi)^2}\int^{\pi/c}_{-\pi/c}dk_z\frac{1}{L_y}\int^{\pi/a}_{-\pi/a}dk_x \frac{e^2\tau E_z}{2\hbar}\frac{\partial f^0(E_-(\mathbf{\tilde{k}}))}{\partial k_z}y_-v_{x, -}(\mathbf{\tilde{k}}),
\end{eqnarray}
where
\begin{math}
\mathbf{\tilde{k}}=(k_x, k_z)
\end{math}
and
\begin{math}
v_{x, \pm}(\mathbf{\tilde{k}})=\frac{1}{\hbar}\frac{\partial E_{\pm}(\mathbf{\tilde{k}})}{\partial k_x}.
\end{math}
Therefore, by noting that $L_y=y_+-y_-$ and $v_{x,+}(\mathbf{\tilde{k}})=-v_{x,-}(\mathbf{\tilde{k}})$, we obtain the KME as
\begin{equation}
M_{z,{\rm slab}}^{\rm KME}=-\frac{e^2\tau E_z}{2\hbar}\frac{1}{(2\pi)^2}\int_{\rm BZ}d\mathbf{\tilde{k}}\frac{\partial f^0(E_-(\mathbf{\tilde{k}}))}{\partial k_z}v_{x, -}(\mathbf{\tilde{k}}).
\label{Mslab1}
\end{equation}
Henceforth we omit the subscript ``$-$'' in $E_-$. In the limit of zero temperature 
\begin{math}
T \rightarrow 0,
\end{math} the $k_z$ derivative of the Fermi-Dirac function is expressed as
\begin{equation}
\frac{\partial f^0(E(\mathbf{\tilde{k}}))}{\partial k_z}=-\delta(E-\mu)\frac{\partial E(\mathbf{\tilde{k}})}{\partial k_z},
\end{equation}
which allows us to rewrite Supplementary Eq.~(\ref{Mslab1}). Thus, in terms of the surface states at the Fermi energy, we can rewrite Supplementary Eq.~(\ref{Mslab1}) as
\begin{eqnarray}
M_{z,{\rm slab}}^{\rm KME}&=&\frac{e^2\tau E_z}{2\hbar^2}\frac{1}{(2\pi)^2}\int^{\pi/c}_{-\pi/c}dk_z\int^{E_u(k_z)}_{E_l(k_z)}dE \delta (E-\mu)\frac{\partial E(k_x, k_z)}{\partial k_z},
\label{Mslab4}
\end{eqnarray}
where $E_l(k_z)$ and $E_u(k_z)$ are the lower and upper bounds of the surface state energies at $k_z$. By performing the integral over $E$, we obtain
\begin{eqnarray}
M_{z,{\rm slab}}^{\rm KME}&=&\frac{e^2\tau E_z}{2\hbar^2}\frac{1}{(2\pi)^2}\int^{\pi/c}_{-\pi/c}dk_z\frac{\partial E(k_x', k_z)}{\partial k_z}\Big|_{E(k_x', k_z)=\mu}.
\label{eq:MIslab}
\end{eqnarray}
This is Eq.~(3) in the main text for Chern insulators, with sgn$\left(\frac{\partial E(k_x', k_z)}{\partial k_x'}\right)=+1$ for the chiral surface states.
We note that the right-hand side of this equation is not identically zero, because the derivative in the integrand is not a total derivative. By noting that
\begin{math}
E(k_x, k_z)=\mu
\end{math}
(constant) yields
\begin{equation}
\frac{\partial E}{\partial k_z}+\frac{d k_x}{d k_z}\Big|_{E=\mu}\frac{\partial E}{\partial k_x}=0,
\end{equation}
we can rewrite Eq.~(\ref{eq:MIslab}) as
\begin{eqnarray}
M_{z,{\rm slab}}^{\rm KME}=-\frac{e^2\tau E_z}{2\hbar}\frac{1}{(2\pi)^2}\int^{\pi/c}_{-\pi/c}dk_zv_x(k_x^{\rm FS}(k_z), k_z)\frac{d k_x^{\rm FS}(k_z)}{d k_z},
\label{Mslab2}
\end{eqnarray}
where $k_x^{\rm FS}$ is a function of $k_z$ defined by
$E(k_x^{\rm FS}(k_z),k_z)=\mu$.

\noindent
{\bf Supplementary Note 2. Extension of the theory to $Z_2$-TIs}

\begin{figure}[t]
\includegraphics[clip,width=0.45\textwidth ]{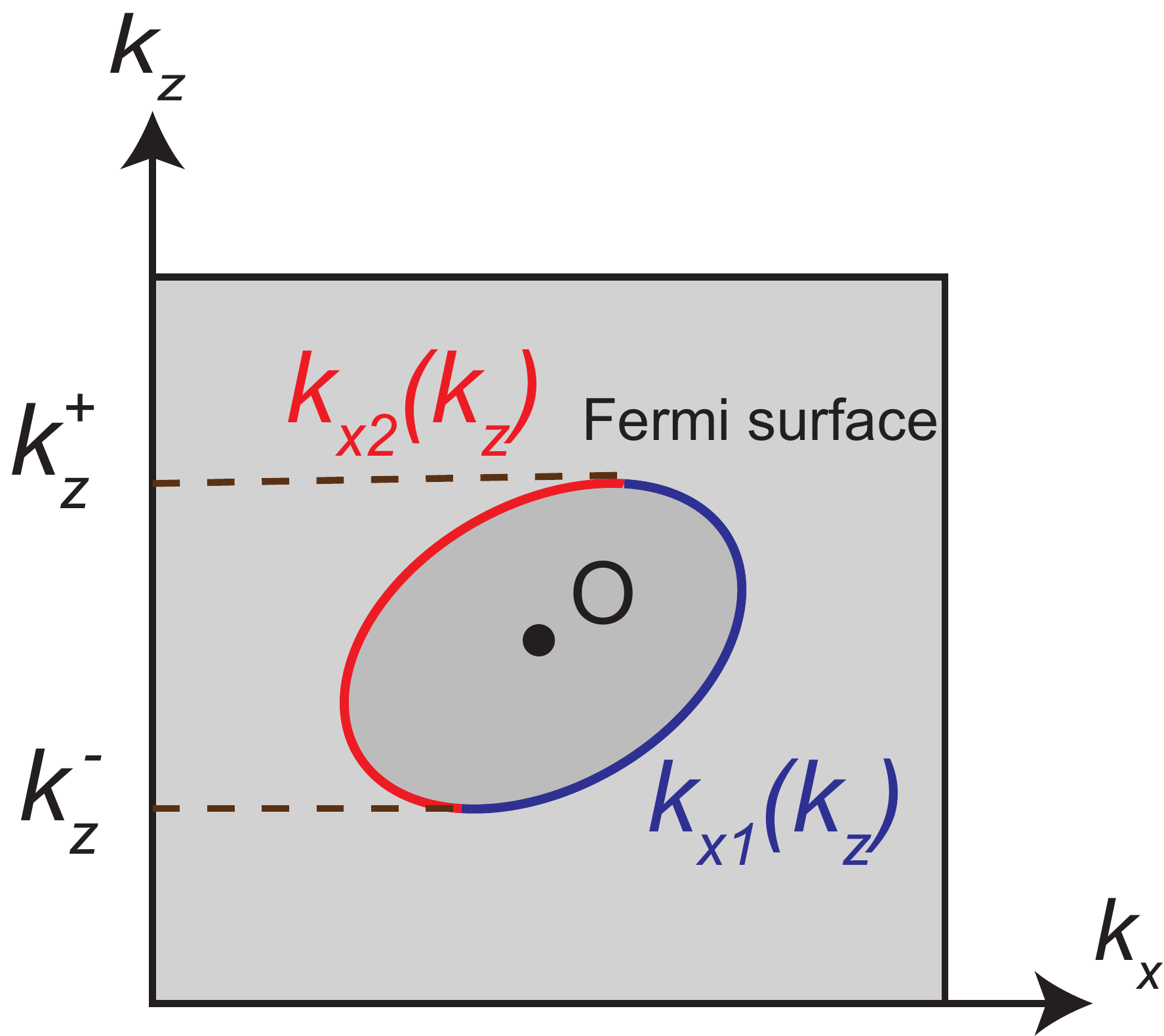} 
\caption{{\bf Fermi surface on the $y=y_-$ surface in the STI.} $k_{x1}$ and $k_{x2}$ are the value of $k_x$ ($k_{x1}>k_{x2}$) on the Fermi surface. Inside and outside the Fermi surface represent the regions of $E_-(k_x, k_z)<\mu$ and $E_-(k_x, k_z)>\mu$, respectively.}
\label{fig:ga}\end{figure}

In Chern insulators, the topological surface states are chiral, and their energy can be a monotonous function of the wavenumber. In contrast, in $Z_2$-TIs, the eigenenergy of the topological surface states is not monotonous. Nonetheless, by dividing the Fermi surface, we can calculate the KME in the same way as in the calculation of Chern insulators. 
For example, in the case of a strong topological insulator (STI) with surface Fermi surfaces given in Fig.~\ref{fig:ga}, the contributions from $k_{x1}(k_z)$ and $k_{x2}(k_z)$ are written from Eq.~(\ref{Mslab4}) as
\begin{eqnarray}
M_{z,{\rm slab},k_{x1}}^{\rm KME}&=&\frac{e^2\tau E_z}{2\hbar}\frac{1}{(2\pi)^2}\int^{k_z^+}_{k_z^-}dk_z\int^{E_b(k_z)}_{E_a(k_z)}dE \delta (E-\mu)\frac{\partial E(k_x', k_z)}{\partial k_z}\Big|_{k_x'=k_{x1}(k_z)}\nonumber\\
&=&\frac{e^2\tau E_z}{2\hbar^2}\int^{\pi/c}_{-\pi/c}\frac{dk_z}{(2\pi)^2}\frac{\partial E(k_x', k_z)}{\partial k_z}\Big|_{k_x'=k_{x1}(k_z)},\label{STIa}\\
M_{z,{\rm slab},k_{x2}}^{\rm KME}&=&\frac{e^2\tau E_z}{2\hbar}\frac{1}{(2\pi)^2}\int^{k_z^+}_{k_z^-}dk_z\int^{E_a(k_z)}_{E_b(k_z)}dE\delta (E-\mu)\frac{\partial E(k_x', k_z)}{\partial k_z}\Big|_{k_x'=k_{x2}(k_z)}\nonumber\\
&=&-\frac{e^2\tau E_z}{2\hbar^2}\int^{\pi/c}_{-\pi/c}\frac{dk_z}{(2\pi)^2}\frac{\partial E(k_x', k_z)}{\partial k_z}\Big|_{k_x'=k_{x2}(k_z)},
\label{STIb}
\end{eqnarray}
where
\begin{eqnarray}
&&E(k_{x1}(k_z), k_z)=\mu,\\
&&E(k_{x2}(k_z), k_z)=\mu,\\
&&E_a(k_z)<\mu<E_b(k_z).
\end{eqnarray}
The sum of Eqs.~(\ref{STIa}) and (\ref{STIb}) is the KME to STIs. Therefore, we obtain the KME in $Z_2$-TIs as the sum of these results:
\begin{eqnarray}
M_{z,{\rm slab}}^{\rm KME}&=&\frac{e^2\tau E_z}{2\hbar^2}\int^{\pi/c}_{-\pi/c}\frac{dk_z}{(2\pi)^2}\frac{\partial E(k_x', k_z)}{\partial k_z}{\rm sgn}\left(\frac{\partial E(k_x', k_z)}{\partial k_x'}\right)\Big|_{E(k_x', k_z)=\mu}.\nonumber
\end{eqnarray}
This formula applies to arbitrary shapes of the surface Fermi surfaces.

Therefore, the magnetoelectric tensor $\alpha_{zz,{\rm slab}}$ for the slab,
defined by $M_{z, {\rm slab}}^{\rm KME}=\alpha_{zz,{\rm slab}}E_z$, is given by
\begin{eqnarray}
\alpha_{zz,{\rm slab}}^{\rm KME}&=&\frac{e^2\tau}{2\hbar^2}\int^{\pi/c}_{-\pi/c}\frac{dk_z}{(2\pi)^2}\frac{\partial E(k_x', k_z)}{\partial k_z}{\rm sgn}\left(\frac{\partial E(k_x', k_z)}{\partial k_x'}\right)\Big|_{E(k_x', k_z)=\mu}.\nonumber\\
\end{eqnarray}
From the above derivation, we can also directly derive that $\alpha_{zz,{\rm slab}}$ is related to the off-diagonal conductivity $\sigma_{xz}$ on the $y=y_-$ surface via
\begin{eqnarray}
\sigma_{zx}=\sigma_{xz}&=&2\alpha_{zz, {\rm slab}}\nonumber \\
&=&\frac{e^2\tau}{\hbar^2}\int^{\pi/c}_{-\pi/c}\frac{dk_z}{(2\pi)^2}\frac{\partial E(k_x', k_z)}{\partial k_z}{\rm sgn}\left(\frac{\partial E(k_x', k_z)}{\partial k_x'}\right)\Big|_{E(k_x', k_z)=\mu}.
\end{eqnarray}
Likewise, the longitudinal conductivity is written as
\begin{eqnarray}
\sigma_{zz}=\frac{e^2\tau}{\hbar^2}\int^{\pi/c}_{-\pi/c}\frac{dk_z}{(2\pi)^2}\frac{\partial E(k_x', k_z)}{\partial k_z}{\rm sgn}\left(\frac{\partial E(k_x', k_z)}{\partial k_z'}\right)\Big|_{E(k_x', k_z)=\mu}.
\end{eqnarray}

\noindent
{\bf Supplementary Note 3. Surface Theory in a one-dimensional prism of 3D Chern insulator}

In this section, for a one-dimensional prism we explain how to rewrite Eq.~(2) into Eq.~(6).
We can determine the eigenstates on the surfaces I and I\hspace{-.1em}I, $u^{\rm I}_{k_xk_z}$ and $u^{\rm I\hspace{-.1em}I}_{k_xk_z}$ from four conditions.
The first condition is that the energy eigenvalues are equal: $E_{k_xk_z}^{\rm I}=E_{k_yk_z}^{\rm I\hspace{-.1em}I}$. The second one is the current conservation at the corner \cite{Raoux2010, concha, takahashi}:
\begin{equation}
v_x^{\rm I}(k_x, k_z)|u^{\rm I}_{k_xk_z}|^2=v_y^{\rm I\hspace{-.1em}I}(k_y, k_z)|u_{k_yk_z}^{\rm I\hspace{-.1em}I}|^2,
\end{equation}
where
\begin{eqnarray}
v_x^{\rm I}(k_x, k_z)=\frac{1}{\hbar}\frac{\partial E^{\rm I}_{k_xk_z}}{\partial k_x},\\
v_y^{\rm I\hspace{-.1em}I}(k_y, k_z)=\frac{1}{\hbar}\frac{\partial E^{\rm I\hspace{-.1em}I}_{k_yk_z}}{\partial k_y}.
\end{eqnarray}
The third one is the periodic boundary condition on the crystal surface:
\begin{equation}
2k_xL_x+2k_yL_y=2\pi n \ \ (n=0,\pm1,\pm2,\dots),
\label{n}
\end{equation}
where $L_x$ and $L_y$ are the lengths of the crystal in the $x$ and $y$ directions. This integer $n$ can be regarded as a quantum number labeling the eigenstates, and we write their eigenvalues as
\begin{math}
E(n, k_z)(=E^{\rm I}_{k_xk_z}=E^{\rm I\hspace{-.1em}I}_{k_yk_z})
\end{math}.
The last one is the normalization condition:
\begin{equation}
2L_x|u_{k_xk_z}^{\rm I}|^2+2L_y|u_{k_yk_z}^{\rm I\hspace{-.1em}I}|^2=1.
\end{equation}
From these conditions, $u_{k_xk_z}^{\rm I}$ and $u_{k_yk_z}^{\rm I\hspace{-.1em}I}$ are
\begin{eqnarray}
|u_{k_xk_z}^{\rm I}|^2=\frac{1}{2v_x^{\rm I}}\frac{1}{\frac{L_x}{v_x^{\rm I}}+\frac{L_y}{v_y^{\rm I\hspace{-.1em}I}}},
|u_{k_yk_z}^{\rm I\hspace{-.1em}I}|^2=\frac{1}{2v_y^{\rm I\hspace{-.1em}I}}\frac{1}{\frac{L_x}{v_x^{\rm I}}+\frac{L_y}{v_y^{\rm I\hspace{-.1em}I}}}.
\label{u}
\end{eqnarray}
Then we obtain KME from Eq.~(3) and Eq.~(\ref{u}):
\begin{eqnarray}
\displaystyle
M_z^{\rm KME}&=&\frac{1}{2\pi}\int^{\pi/c}_{-\pi/c}dk_z\sum_n\frac{e\tau E_z}{\hbar}\frac{\partial f^0(E(n, k_z))}{\partial k_z}\left(-\frac{e}{2}\right)\frac{1}{\frac{L_x}{v_x^{\rm I}}+\frac{L_y}{v_y^{\rm I\hspace{-.1em}I}}}.
\end{eqnarray}

When $L_x$ and $L_y$ are large, $k_x$, $k_y$ and $E(n, k_z)$ are regarded as continuous variables.  By rewriting the summation over $n$ to an integral over $E$ and by using Eq.~(\ref{n}), we obtain the KME as
\begin{eqnarray}
\displaystyle
M_z^{\rm KME}&=&-\frac{e^2\tau E_z}{2\hbar}\frac{1}{2\pi}\int^{\pi/c}_{-\pi/c}\int dEdk_z\frac{\partial f^0(E(n, k_z))}{\partial k_z}\frac{dn}{dE}\frac{1}{\frac{L_x}{v_x^{\rm I}}+\frac{L_y}{v_y^{\rm I\hspace{-.1em}I}}}\nonumber\\
&=&\frac{e^2\tau E_z}{\hbar^2}\frac{1}{(2\pi)^2}\int^{\pi/c}_{-\pi/c}dk_z\frac{\partial E(n, k_z)}{\partial k_z}\big|_{E(n, k_x)=\mu},
\label{Mone3}
\end{eqnarray}
where we used
\begin{eqnarray}
&&\frac{dn}{dE}=\frac{1}{\hbar\pi}\left(\frac{L_x}{v_x^{\rm I}}+\frac{L_y}{v_y^{\rm I\hspace{-.1em}I}}\right),\\
&&\frac{\partial f^0(E(n, k_z))}{\partial k_z}=-\delta(E-\mu)\frac{\partial E(n, k_z)}{\partial k_z}
\end{eqnarray}
From Eq.~(\ref{n}), let $k_x^{{\rm I}0}(k_z,E)$ and $k_y^{{\rm I\hspace{-.1em}I}0}(k_z,E)$ be the values of $k_x$ and $k_y$ for a given value of $k_z$ and $E$. Therefore they satisfy
\begin{equation}
n=\frac{1}{\pi}(L_xk^{{\rm I}0}_x(k_z, E)+L_yk^{{\rm I\hspace{-.1em}I}0}_y(k_z, E))
\end{equation}
Therefore, we finally get:
\begin{eqnarray}
&&\frac{\partial E(n, k_z)}{\partial k_z}+\frac{\partial E(n, k_z)}{\partial n}\frac{1}{\pi}\left(L_x\frac{\partial k_x^{{\rm I}0}(k_z,E)}{\partial k_z}+L_y\frac{\partial k_y^{{\rm I\hspace{-.1em}I}0}(k_z,E)}{\partial k_z}\right)=0,\nonumber\\
\end{eqnarray}
Therefore, we rewrite Eq.~(\ref{Mone3}) as
\begin{equation}
M_z^{\rm KME}=-\frac{e^2\tau E_z}{\hbar}\frac{1}{(2\pi)^2}\int^{\pi/c}_{-\pi/c}dk_z\frac{L_x\frac{\partial k_x^{{\rm I}}}{\partial k_z}+L_y\frac{\partial k_y^{{\rm I\hspace{-.1em}I}}}{\partial k_z}}{\frac{L_x}{v_x^{\rm I}}+\frac{L_y}{v_y^{\rm I\hspace{-.1em}I}}}\Big|_{E=\mu},
\label{Mone2}
\end{equation}
where $v_x^{\rm I}=\frac{1}{\hbar}\frac{\partial E^{\rm I}}{\partial k_{x}}$ and $v_y^{\rm I\hspace{-.1em}I}=\frac{1}{\hbar}\frac{\partial E^{\rm I\hspace{-.1em}I}}{\partial k_{y}}$ and
\begin{math}k_x^{{\rm I}}(k_z, E)
\end{math}
and 
\begin{math}
k_y^{{\rm I\hspace{-.1em}I}}(k_z, E)
\end{math}
are functions obtained from $E=E^{\rm I}_{k_xk_z}$ and $E=E^{\rm I\hspace{-.1em}I}_{k_yk_z}$, respectively.
This is the formula for KME in a one-dimensional prism of a three-dimensional Chern insulator.

\noindent
{\bf Supplementary Note 4. KME and electric conductivity tensors}

We can also derive KME in a Chern insulating system by using electric conductivity tensors on surfaces.
 On the surface I along the $xz$ plane, the surface current density along $xz$ plane, $j_x$ and $j_z$ in response to the electric field $(E_x, E_z)$, is represented as
\begin{eqnarray}
j_x^{\rm I}&=&\sigma_{xx}^{\rm I}E_x^{\rm I}+\sigma_{xz}^{\rm I}E_z, 
\label{start}\\
j_z^{\rm I}&=&\sigma_{zx}^{\rm I}E_x^{\rm I}+\sigma_{zz}^{\rm I}E_z.
\end{eqnarray}
where $\mathbf{\sigma}^{\rm I}_{ij}$ is the electric conductivity tensor. Since we assumed that the system is invariant under twofold rotation with respect to the $z$-axis, the surface I\hspace{-.1em}I\hspace{-.1em}I has the same tranport property with the surface I, but with an opposite normal vector. Namely, it has
\begin{eqnarray}
j_x^{\rm I\hspace{-.1em}I\hspace{-.1em}I}&=&\sigma_{xx}^{\rm I}E_x^{\rm I\hspace{-.1em}I\hspace{-.1em}I}-\sigma_{xz}^{\rm I}E_z,\\
j_z^{\rm I\hspace{-.1em}I\hspace{-.1em}I}&=&-\sigma_{zx}^{\rm I}E_x^{\rm I\hspace{-.1em}I\hspace{-.1em}I}+\sigma_{zz}^{\rm I}E_z.
\end{eqnarray}
We note that the electric field along the $z$-axis, $E_z$, is common among the four surfaces. Similarly, on the surface I\hspace{-.1em}I along the $yz$ plane, the surface current density along the $yz$ plane, $j_y$ and $j_z$ in response the electric field $(E_y, E_z)$, is written as
\begin{eqnarray}
j_y^{\rm I\hspace{-.1em}I}&=&\sigma_{yy}^{\rm I\hspace{-.1em}I}E_y^{\rm I\hspace{-.1em}I}+\sigma_{yz}^{\rm I\hspace{-.1em}I}E_z,\\
j_z^{\rm I\hspace{-.1em}I}&=&\sigma_{zy}^{\rm I\hspace{-.1em}I}E_y^{\rm I\hspace{-.1em}I}+\sigma_{zz}^{\rm I\hspace{-.1em}I}E_z.
\end{eqnarray}
Similarly, on the surface I\hspace{-.1em}V one has has
\begin{eqnarray}
j_y^{\rm I\hspace{-.1em}V}&=&\sigma_{yy}^{\rm I\hspace{-.1em}I}E_y^{\rm I\hspace{-.1em}V}-\sigma_{yz}^{\rm I\hspace{-.1em}I}E_z,\\
j_z^{\rm I\hspace{-.1em}V}&=&-\sigma_{zy}^{\rm I\hspace{-.1em}I}E_y^{\rm I\hspace{-.1em}V}+\sigma_{zz}^{\rm I\hspace{-.1em}I}E_z.
\end{eqnarray}
We note that the electric field satisfies
\begin{eqnarray}
\oint_{\rm surface}\mathbf{E}\cdot d\mathbf{l}&=&L_x(E_x^{\rm I}-E_x^{\rm I\hspace{-.1em}I\hspace{-.1em}I})+L_y(E_y^{\rm I\hspace{-.1em}I}-E_y^{\rm I\hspace{-.1em}V})=0,
\end{eqnarray}
and that the current around the crystal is conserved:
\begin{eqnarray}
j_{\rm circ}\equiv j_x^{\rm I}=j_y^{\rm I\hspace{-.1em}I}=-j_x^{\rm I\hspace{-.1em}I\hspace{-.1em}I}=-j_y^{\rm I\hspace{-.1em}V}.
\label{end}
\end{eqnarray}
Combining Eqs.~(\ref{start})-(\ref{end}), we obtain
\begin{equation}
M_z^{\rm KME}=j_{\rm circ}=\frac{\frac{L_x}{\sigma_{xx}^{\rm I}}\sigma_{xz}^{\rm I}+\frac{L_y}{\sigma_{yy}^{\rm I\hspace{-.1em}I}}\sigma_{yz}^{\rm I\hspace{-.1em}I}}{\frac{L_x}{\sigma_{xx}^{\rm I}}+\frac{L_y}{\sigma_{yy}^{\rm I\hspace{-.1em}I}}}E_z.\nonumber
\end{equation}

\noindent
{\bf Supplementary Note 5. Finite-size effect}

In this section, we introduce our fitting result in detail. We calculate the KME with various system sizes from Eq.~(2). By using these results, we determine the parameters in the fitting function (see Methods). With the hopping parameters $t_x=b_x=1.1t_y$, $t_y=b_y=m$, $t_4=0.15t_y$, $t_3=0.001t_y$ and $\mu=0$, we get the results fitting as
\begin{equation}
\frac{M_z^{\rm KME}}{K}=\frac{0.0082L_x+2.62L_y-1.47-\frac{1.51}{L_x}-\frac{-1.1}{L_y}}{L_x+1.17L_y}\times10^{-5},
\end{equation}
where 
\begin{math}
K\equiv e^2\tau E_z t_y/\hbar^2
\end{math}
, if we take the limit of $L_x\to\infty$ and $L_y\to\infty$ with the ratio $L_x/L_y$ kept constant, we get:
\begin{equation}
\frac{M_z^{\rm KME}}{K}=\frac{0.0082L_x+2.62L_y}{L_x+1.17L_y}\times10^{-5}.
\label{eq:MK1}
\end{equation}
 On the other hand, the result of Eq.~(7) based on the surface theory is fitted as:
\begin{equation}
\frac{M_z^{\rm KME}}{K}=\frac{0.0210L_x+2.46L_y}{L_x+1.10L_y}\times10^{-5}.
\label{eq:MK2}\end{equation}
Apart from the coefficients of $L_x$ in the numerators, Eqs.~(\ref{eq:MK1}) and (\ref{eq:MK2}) agree well. Since the effect of $L_x$ in the numerators of Egs.~(\ref{eq:MK1}) and (\ref{eq:MK2}) is much smaller than that of $L_y$, it is difficult to numerically obtain the coefficient of $L_x$ in the numerator with good accuracy. Thus, we conclude that these results agree well as shown in Fig.~4d.

\noindent
{\bf Supplementary Note 6. Cu$_2$ZnGeTe$_4$}

After searching in the topological material database \cite{zhang2019catalogue}, we notice that Cu$_2$ZnSnSe$_4$ \cite{guen1979physical} with $\bm{{\it 82}}$ and CdGeAs$_{2}$ with $\bm{{\it 122}}$ are two ideal candidates of $Z_2$-TIs with a direct gap
. Although Cu$_2$ZnGeTe$_4$ \cite{guen1979physical} is also classified to be a $Z_2$-TI according to the topological material database, there are several electron and hole pockets at the Fermi energy.

\noindent
{\bf Supplementary Note 7. Magnetoelectric tensor for the space group $\bm{{\it 82}}$}

In the main text, the  kinetic magnetoelectric tensor for the space group $\bm{{\it 82}}$ ($S_4$), defined by 
$\bm{M}=\bm{\alpha}\bm{E}$, is
\begin{eqnarray}
\alpha_{\bm{{\it 82}}}= \left[
 \begin{matrix}
   \alpha_{11} & \alpha_{12} & 0 \\
   \alpha_{12} & -\alpha_{11} & 0 \\
   0 & 0 & 0
  \end{matrix}  \right].
\label{tensor}
\end{eqnarray}
As an example, we calculate $\alpha_{33}$ in a one-dimensional quadrangular prism of Cu$_2$ZnSnSe$_4$ from Eq.~(8). From $S_4$ symmetry in  Cu$_2$ZnSnSe$_4$, we get $\sigma_{11}^{\rm I}=\sigma_{22}^{\rm I\hspace{-.1em}I}$ and $\sigma_{13}^{\rm I}=-\sigma_{23}^{\rm I\hspace{-.1em}I}$. Hence, when the quadrangular prism preserves the $S_4$ symmetry, i.e. when $L_x=L_y$, we get $\alpha_{33}=0$ from Eq.~(8), in accordance with Eq.~(\ref{tensor}). On the other hand, when $L_1\neq L_2$, $\alpha_{33}$ become nonzero as opposed to Eq.~(\ref{tensor}), because the $S_4$ symmetry is broken by the shape of the sample. 

Thus, in general systems, if a component of the kinetic magnetoelectric tensor $\alpha_{ij}$ is nonzero from symmetry analysis, it means that the corresponding KME appears regardless of the crystal shape, reflecting the symmetry of the crystal. On the other hand, when $\alpha_{ij}=0$, the KME does not occur when the crystal shape preserve the point-group symmetry of the crystal, but otherwise the KME may occur depending on the crystal shape.

\noindent
{\bf Supplementary Note 8. [010] surface states of Cu$_2$ZnSnSe$_4$}

\begin{figure}[t]
\raisebox{-\height}{\includegraphics[width=0.58\textwidth]{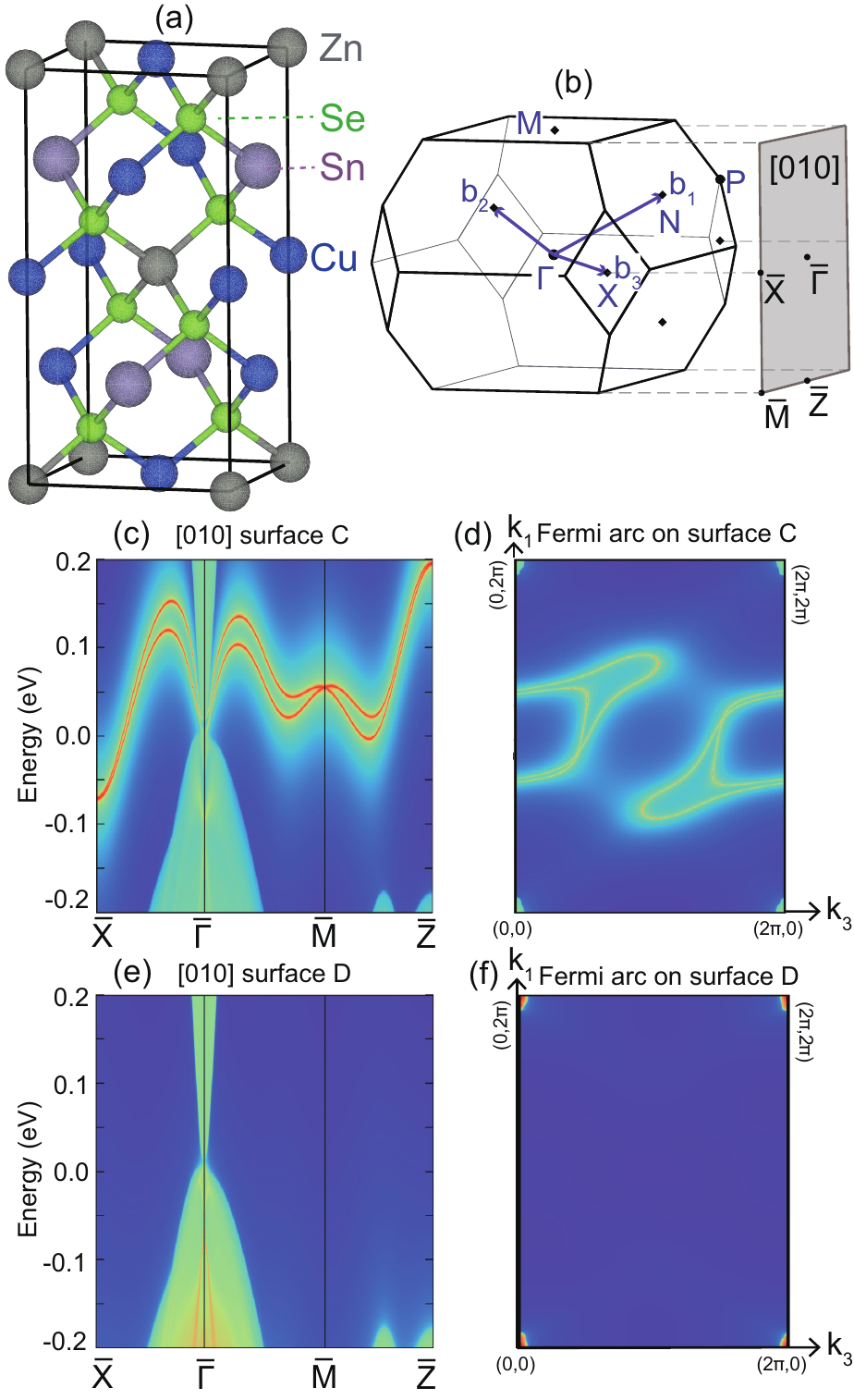}}
\caption{{\bf First principle calculations of Cu$_2$ZnSnSe$_4$ on the [010] surface.} ({\bf a-b}) Crystal structure and Brillouin zone for the bulk and surface along [010] direction. ({\bf c-d}) Surface states and Fermi arcs calculation on the [010] surface with Cu-Sn layer termination  (surface C). ({\bf e-f}) Surface states and Fermi surface calculation on the [010] surface with Se layer termination  (surface D), which has a very small Fermi surface buried in the bulk states. }
\label{SImat}
\end{figure}

In the main text we calculated the [001] surface of Cu$_2$ZnSnSe$_4$. 
In this Supplementary note we calculate the surface of [010] direction, we also calculated the surface states and magnetoelectric susceptibility with two different terminations, i.e., surface C with Cu-Sn layer termination and surface D with Se layer termination, as shown in Fig.~\ref{SImat}. On the surface D, which is the Se layer termination, the Fermi arc is very small and buried in the bulk states, so we only calculate the $\alpha_{11}^{C}$ for illustration. On the surface C, there is big hole-like pocket formed by the Dirac cone at $\bar{X}$, which contributes to large magnetoelectric susceptibility of $\alpha_{11}^{\rm C}=-2.324\times 10^{9}\tau s^{-1}\Omega^{-1}$.

\clearpage


\noindent
{\bf Supplementary Note 9. Surface theory of the spin Edelstein effect in a slab of a 3D topological insulator}

\begin{figure}[t]
\includegraphics[clip,width=0.8\textwidth]{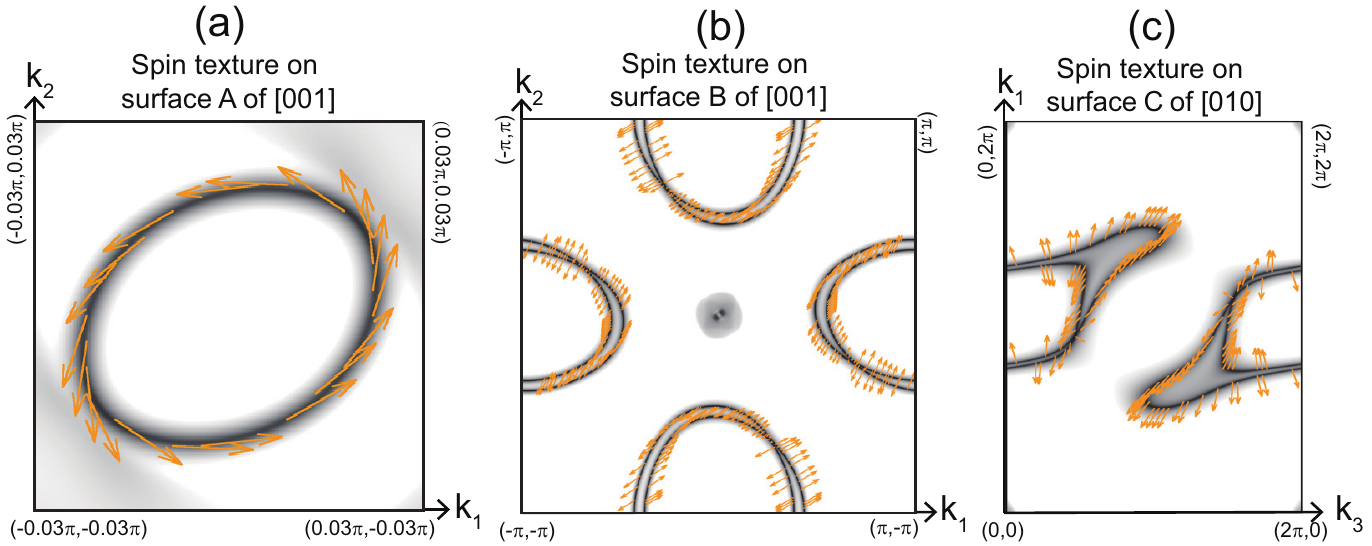}
\caption{{\bf Spin textures of the surface states of the 
 $Z_2$ TI Cu$_2$ZnSnSe$_4$.} The surface states are shown for the two surface terminations A and B for the [001] surface and one termination C for the [010] surface, shown in (a), (b) and (c), respectively. The black curves represent the surface Fermi surfaces, and the orange arrows represent the spin polarization for the surfae states.}
 \label{fig:spin}
\end{figure}

In this section, we explain how to calculate the spin Edelstein effect in a slab of a 3D topological insulator, in order to compare its behavior with the orbital analogue, 
the kinetic magnetoelectric effect.
Let us set the 3D topological insulator in a slab geometry, which is infinite along the $xz$ plane and has a
finite width $L_y$ along the $y$ axis.
Let
\begin{math}
\psi_{+(-)}(k_x, k_z)
\end{math}
denote a wave function of a surface state on the $y=y_+$ $(y=y_-)$ surface of the crystal. 
From Eq.~(3), we evaluate the spin Edelstein effect as
\begin{eqnarray}
\displaystyle
M_{z, {\rm slab}}^{\rm KME}&=&M_{z, +}+M_{z, -},\\
M_{z, +}&=&-\frac{1}{(2\pi)^2}\int^{\pi/c}_{-\pi/c}dk_z\frac{1}{L_y}\int^{\pi/a}_{-\pi/a}dk_x \frac{e\mu_B\tau E_z}{\hbar}\frac{\partial f^0(E_+(\mathbf{\tilde{k}}))}{\partial k_z}s_{z, +}(\mathbf{\tilde{k}}),\\
M_{z, -}&=&-\frac{1}{(2\pi)^2}\int^{\pi/c}_{-\pi/c}dk_z\frac{1}{L_y}\int^{\pi/a}_{-\pi/a}dk_x \frac{e\mu_B\tau E_z}{\hbar}\frac{\partial f^0(E_-(\mathbf{\tilde{k}}))}{\partial k_z}s_{z, -}(\mathbf{\tilde{k}}),
\end{eqnarray}
where
$\mu_B$ is the Bohr magneton, 
\begin{math}
\mathbf{\tilde{k}}=(k_x, k_z)
\end{math}
and
\begin{math}
s_{z, \pm}(\mathbf{\tilde{k}})=\psi_{\pm}^{\dagger}s_z
\psi_{\pm}
\end{math}
with $s_z$ being the Pauli matrix.
We here assume twofold rotational symmetry $C_{2z}$ around the $z$ axis. Then it leads to $s_{z,+}(\mathbf{\tilde{k}})=s_{z,-}(\mathbf{\tilde{k}})$, and we obtain the spin Edelstein effect as
\begin{equation}
M_{z,{\rm slab}}^{\rm KME}=-\frac{e\mu_B\tau E_z}{\hbar L_y}\frac{1}{2\pi^2}\int_{\rm BZ}d\mathbf{\tilde{k}}\frac{\partial f^0(E_-(\mathbf{\tilde{k}}))}{\partial k_z}s_{z, -}(\mathbf{\tilde{k}}).
\label{spinMslab1}
\end{equation}
Henceforth we omit the subscript ``$-$'' in $E_-$. In the limit of zero temperature 
\begin{math}
T \rightarrow 0,
\end{math} the $k_z$ derivative of the Fermi-Dirac function is expressed as
\begin{equation}
\frac{\partial f^0(E(\mathbf{\tilde{k}}))}{\partial k_z}=-\delta(E-\mu)\frac{\partial E(\mathbf{\tilde{k}})}{\partial k_z}.
\end{equation}
Therefore, in terms of the surface states at the Fermi energy, we can rewrite Eq.~(\ref{spinMslab1}) as
\begin{eqnarray}
M_{z,{\rm slab}}^{\rm KME}&=&\frac{e\mu_B\tau E_z}{\hbar L_y}\frac{1}{2\pi^2}
\int_{\rm BZ}d\mathbf{\tilde{k}}\delta (E-\mu)\frac{\partial E(k_x, k_z)}{\partial k_z}s_{z}(\mathbf{\tilde{k}}),
\label{spinMslab4}
\end{eqnarray}
This $\delta$-function can be rewritten as a function of $k_z$ as
\begin{equation}
\delta(E-\mu)=\sum_n\frac{\delta(k_z-k_z^{(n)})}{
\left|\frac{\partial E(k_z, k_z)}{\partial k_z}\right|},
\end{equation}
where the sum is taken over $k_z^{(n)}$ satisfying $E(k_z^{(n)})=\mu$ for a given $k_x$.
Thus we get
\begin{eqnarray}
M_{z,{\rm slab}}^{\rm KME}&=&\frac{e\mu_B\tau E_z}{\hbar L_y}\frac{1}{2\pi^2}
\int^{\pi/a}_{-\pi/a}dk_x
\text{sgn}\left(\frac{\partial E(k_x, k_z)}{\partial k_z}\right)_{E=\mu}
s_{z}(\mathbf{\tilde{k}}).
\label{spinMslab5}
\end{eqnarray}
Using this equation, we calculate the spin magnetization induced by the 
electric field in the $Z_2$ TI Cu$_2$ZnSnSe$_4$. From our first-principle calculation in Fig.~\ref{fig:spin}, we obtain $\alpha_{11}^{\text{A,spin}}=2.798\times10^{-2} \text{ms}^{-1}\Omega^{-1}$ $\cdot \tau/L_y$, 
$\alpha_{11}^{\text{B,spin}}=-2.575 \text{ms}^{-1}\Omega^{-1}$ $\cdot \tau/L_y$, and
$\alpha_{11}^{\text{C,spin}}=-7.818 \text{ms}^{-1}\Omega^{-1}$ $\cdot \tau/L_y$.
Thus it inversely scales with the system size along the $y$ direction.
By dividing them by the sheet resistivities, they lead to 
$\alpha_{11}^{\text{A,spin}}/\sigma_{11,\square}^{\rm A}=3.8\times10^{-11}{\rm m}/L_y$
and
$\alpha_{11}^{\text{B,spin}}/\sigma_{11,\square}^{\rm B}=1.3\times10^{-10}{\rm m}/L_y$.
For example, with the sample with $L_y=1$mm, they are 
$\alpha_{11}^{\text{A,spin}}/\sigma_{11,\square}^{\rm A}=3.8\times10^{-8}$
and
$\alpha_{11}^{\text{B,spin}}/\sigma_{11,\square}^{\rm B}=1.3\times10^{-7}$.
Compared with these values for the  spin Edelstein effect,  the orbital counterpart, i.e. the kinetic 
magnetoelectric effect, has $\alpha_{11,\square}^{\text{A}}/\sigma_{11}^{\rm A}=0.25$
and
$\alpha_{11}^{\text{B}}/\sigma_{11,\square}^{\rm B}=0.13$, which are much larger 
by 6 orders of magnitude.

\newpage

\noindent
{\bf Supplementary References}